\documentclass[11pt,a4paper,twoside,groupcitations]{article}
\usepackage[T1]{fontenc}
\usepackage[ansinew]{inputenc}
\usepackage[english]{babel}
\usepackage{amsfonts}
\usepackage{amsmath}
\usepackage{bm}
\usepackage{array}
\usepackage{amsthm}
\usepackage{amssymb}
\usepackage{graphicx}
\usepackage{subfigure}
\usepackage{braket}
\usepackage{eucal}
\usepackage{verbatim}
\usepackage[table]{xcolor}
\usepackage{caption}
\usepackage{cite}
\usepackage{textcomp}
\usepackage{multicol}
\usepackage{tikz}
\usetikzlibrary{positioning,arrows}
\usetikzlibrary{decorations.pathmorphing}
\usetikzlibrary{decorations.markings}
\usetikzlibrary{calc,decorations.markings}
\usetikzlibrary{arrows,shapes}
\usetikzlibrary{matrix,arrows}
\usepackage{pgfplots}
\usepackage{xparse}
\definecolor{jade}{HTML}{00A86B}
\newcommand{\be}{\begin{eqnarray}}
\newcommand{\ee}{\end{eqnarray}}

\newcommand{\expec}[1]{\mbox{$\langle\, #1\,\rangle$}}

\renewcommand{\a}{\hat a}
\newcommand{\ac}{\hat a^{\dagger}}

\renewcommand{\d}{\mbox{${\rm d}$}} %d differenziale non corsivo in math mode
\newcommand{\lp}{\ell_{\rm p}}
\newcommand{\mpl}{m_{\rm p}}
\newcommand{\gn}{G_{\rm N}}
\newcommand{\rh}{r_{\rm H}}

\newcommand{\Ng}{{N_{\rm G}}}

\newcommand{\dd }{{\mathrm d}}
\DeclareMathOperator{\Si}{Si}
\title{\bf Quantum Reissner-Nordstr\"om geometry: singularity and Cauchy horizon}
\author{Roberto~Casadio$^{ab}$\thanks{E-mail: casadio@bo.infn.it},
Andrea~Giusti$^{c}$\thanks{E-mail: agiusti@phys.ethz.ch}
$\ $and
Jorge~Ovalle$^{de}$\thanks{E-mail: jorge.ovalle@physics.slu.cz}
\\
\\
$^a${\em Dipartimento di Fisica e Astronomia, Universit\`a di Bologna}
\\
{\em via Irnerio~46, 40126 Bologna, Italy}
\\
\\
$^b${\em I.N.F.N., Sezione di Bologna, I.S.~FLAG}
\\
{\em viale B.~Pichat~6/2, 40127 Bologna, Italy}
\\
\\
$^c$ {\em Institute for Theoretical Physics, ETH Zurich}
\\
{\em Wolfgang-Pauli-Strasse 27, 8093 Zurich, Switzerland}
\\
\\
$^d$ {\em Research Centre of Theoretical Physics and Astrophysics}
\\
{\em 	Institute of Physics, Silesian University in Opava, CZ-746 01 Opava, Czech Republic}
\\
\\
$^d$ {\em Departamento de F\'isica, Facultad de Ciencias B\'asicas,}
\\
{\em 	Universidad de Antofagasta, Chile.}
}
\begin{document}
\maketitle
\begin{abstract}
We present a quantum description of electrically charged spherically symmetric black holes 
given by coherent states of gravitons in which both the central singularity and the Cauchy 
horizon are not realised.
\end{abstract}
\section{Introduction and motivation}
\setcounter{equation}{0}
\label{Sintro}
Classical black hole solutions of general relativity contain spacetime singularities~\cite{HE},
which are expected to be removed in the quantum theory of the gravitational collapse of
a compact source (see, {\em e.g.}, Refs.~\cite{OS}).
Moreover, charged and rotating classical black holes also contain a inner Cauchy horizon,
which signals a potential loss of predictability and also gives rise to mass inflation
at the perturbative level (see, {\em e.g.}, Refs.~\cite{CH}).
These latter considerations are the underlying motivations for the
{\em strong cosmic censorship\/} conjecture~\cite{Penrose:1,Penrose:2},
which can be simply phrased as the fact that the evolution of some sufficiently regular
initial data should always give rise to a globally hyperbolic spacetime~\cite{Brady:1998au}.
In particular, this conjecture implies that perturbations of the inner Cauchy horizon should
turn it into a curvature singularity.
\par
Quite interestingly, many candidates as regular black holes appearing in the literature
(for an incomplete list, see Refs.~\cite{regular}) also display a inner horizon 
(with some interesting exceptions, like those in Refs.~\cite{regularH}).
One might then wonder if trading the central singularity for a Cauchy horizon represents
a real progress in our understanding of black hole physics.
More generally, one would like to understand how regular the geometry really needs to be
for physical consistency and whether this trade-off can be avoided.
In this respect, it is also interesting to note that the inner horizon that forms in models of
gravitational collapse~\cite{OS} is not eternal.
Therefore, a regular black hole with an inner horizon generated by the gravitational collapse
does not necessarily have (lasting) issues concerning the initial value problem, 
or the instabilities associated with Cauchy horizons. 
On the other hand, there is no guarantee that these issues are always
avoided in collapse models and one must check on a case by case basis.
\par
In Ref.~\cite{Casadio:2021eio}, a quantum coherent state for the spherically symmetric
and electrically neutral Schwarzschild metric was introduced, following ideas from
Refs.~\cite{Casadio:2016zpl,Casadio:2017cdv,Barnich:2010bu,Mueck:2013mha}.
This coherent state is built for a scalar field, which in turn is meant to effectively describe
the geometry itself as a gravitational potential emerging from the (longitudinal or temporal)
polarisations of the graviton.
It was in particular shown in Ref.~\cite{Casadio:2021eio} that the conditions for the very existence
of such a quantum state require departures from the purely classical behaviour both in the
infrared (IR) and, more importantly, in the ultraviolet (UV).
The IR behaviour can be connected with the finite extent of our causally connected
Universe~\cite{Giusti:2021shf}.
The UV deviation from the classical general relativistic vacuum can instead
be interpreted as the existence of a (quantum) extended matter core,
which sources the geometry~\cite{Casadio:2021cbv} and gives rise
to quantum hair~\cite{qhair,Ovalle:2020kpd}.
This quantum core indeed removes the central singularity and keeps tidal forces
everywhere finite.
Geometrical quantities like the Ricci curvature and the Kretschmann scalar 
still diverge towards the origin, but their integrals remain finite, which corresponds
to having a so-called integrable singularity.
Furthermore, no inner horizon appears for any size of the matter core.
\par
In order to investigate the causal structure of spherically symmetric quantum black holes
with electric charge, in this work we apply the approach of Ref.~\cite{Casadio:2021eio}
to the Reissner-Nordstr\"om metric
\be
\d s^2
=
-\left(1+2\,V_{\rm RN}\right)\d t^2
+
\frac{\d r^2}{1+2\,V_{\rm RN}}
+
r^2\,\d\Omega^2
\ ,
\label{gRN}
\ee
with~\footnote{We shall use units with $c=1/4\,\pi\,\epsilon_0=1$, $\gn=\lp/\mpl$,
$\hbar=\lp\,\mpl$, with $\lp$ the Planck length and $\mpl$ the Planck mass.
Hence, the combination $\gn\,Q^2$ has dimensions of a length squared.}
\be
V_{\rm RN}
=
-\frac{\gn\,M}{r}
+
\frac{\gn\,Q^2}{2\,r^2}
\equiv
V_{M}
+
V_{Q}
\ ,
\label{Vrn}
\ee
where $M$ is the ADM mass~\cite{adm} and $Q$ the charge of the black hole.
We recall that, for $\gn\,M^2>Q^2$, the above spacetime contains two horizons
determined by $g^{rr}=1+2\,V_{\rm RN}=0$, namely
\be
R_\pm
=
\gn\,M
\pm
\sqrt{\gn^2\,M^2-\gn\,Q^2}
\ ,
\label{R+-}
\ee
with $R_+$ being the event horizon and $R_-$ a Cauchy horizon.
The quantum version of the functions $V_M$ and $V_Q$ in Eq.~\eqref{Vrn} will be employed
in order to reconstruct a quantum corrected complete metric to replace~\eqref{gRN}.
From this new metric, one can then analyse how the necessary material core predicted
by quantum physics affects the singularity, inner horizon and thermodynamics.
In particular, we assume that it is in this core that resides the charge $Q$.
\par
In Section~\ref{s:coherent}, we will derive in details the coherent state for the
metric~\eqref{gRN} and the corresponding quantum metric will then be analysed
in Section~\ref{s:qmetric};
concluding remarks and outlooks will be provided in Section~\ref{s:conc}.
\section{Quantum coherent state for the Reissner-Nordstr\"om geometry}
\label{s:coherent}
\setcounter{equation}{0}
Following Ref.~\cite{Casadio:2021eio}, we assume that the quantum vacuum $\ket{0}$ corresponds
to a spacetime devoid of any matter or gravitational excitations.
In order to effectively describe the gravitational excitations giving rise to the geometry,
we first rescale the potential $V_{\rm RN}$ so as to obtain a canonically normalised real scalar field
$\Phi =\sqrt{{\mpl}/{\lp}}\, V_{\rm RN}$, and then quantise $\Phi$ as a massless field satisfying
the free wave equation in the Minkowski spacetime
\be
\left[
-\frac{\partial^2}{\partial t^2}
+
\frac{1}{r^2}\,\frac{\partial}{\partial r}
\left(r^2\,\frac{\partial}{\partial r}\right)
\right]
\Phi(t,r)
=
0
\ .
\label{KG}
\ee
Since we are only interested in static configurations, it is convenient to choose the normal modes of
Eq.~\eqref{KG} described in terms of spherical Bessel functions $j_0={\sin(k\,r)}/{k\,r}$ for
$k>0$, that is
\be
u_{k}(t,r) = e^{-i\,k\,t}\,j_0(k\,r)
\ .
\ee
The quantum field operator,
\be
\label{Phi}
\hat{\Phi}(t,r)
=
\int\limits_0^\infty
\frac{k^2\,\dd k}{2\,\pi^{2}}\,
\sqrt{\frac{\hbar}{2\,k}}
\left[
\a_{k}\,
u_k(t,r)
+
\ac_{k}\, 
u^*_k(t,r)
\right]
\ ,
\ee
and its conjugate momentum,
\be
\hat{\Pi}(t,r) 
=
i\int\limits_0^\infty
\frac{k^2\,\dd k}{2\,\pi^{2}}\,
\sqrt{\frac{\hbar\,k}{2}}
\left[
\a_{k}\,
u_k(t,r)
-
\ac_{k}\, 
u^*_k(t,r)
\right]
\ ,
\ee
satisfy the equal time commutation relations,
\be
\left[\hat{\Phi}(t,r)
,\hat{\Pi}(t,s)\right] 
=
\frac{i\,\hbar}{4\,\pi\,r^2}\,
\delta(r-s)
\ ,
\ee
if
\be
\left[\a_{k},\ac_{p}\right]
=
\frac{2\,\pi^2}{k^2}\,
\delta(k-p)
\ .
\ee
The Fock space is then built from the vacuum defined by $\a_{k}\ket{0}=0$
for all $k>0$.
\par
The classical static configurations~\eqref{Vrn} are realised in the quantum theory
as coherent states $\ket{g}$ such that $\a_{k} \ket{g} = g(k)\,e^{i\,\gamma_{k}(t)} \ket{g}$
and
\be
\sqrt{\frac{\lp}{\mpl}}
\bra{g}\hat{\Phi}(t,r)\ket{g}
=
V_{\rm RN}(r)
\ .
\label{expecphi}
\ee
From the expansion~\eqref{Phi}, we obtain
\be
\bra{g}\hat{\Phi}(t,r)\ket{g}
=
\int\limits_0^\infty
\frac{k^2\,\dd k}{2\,\pi^2}\,
\sqrt{\frac{2\,\lp\,\mpl}{k}}\,
g(k)\,
\cos[\gamma_k(t)-k\,t]\,
j_0(k\,r)
\ ,
\ee
and time-independence is obtained by setting $\gamma_k=k\,t$.~\footnote{We recall that
static potentials are obtained from non-propagating modes in quantum field
theory~\cite{Feynman,Barnich:2010bu,Mueck:2013mha}.}
If we write the classical potential as
\be
V_{\rm RN}
=
\int\limits_0^\infty
\frac{k^2\,\dd k}{2\,\pi^2}\,
\left[\tilde V_M(k)+\tilde V_Q(k)\right]
j_0(k\,r)
\ ,
\label{Vk}
\ee
we obtain
\be
\tilde{V}_M
=
-4\,\pi\,\gn\,\frac{M}{k^2}
\ee
and 
\be
\tilde{V}_Q
=
\pi^2\,\gn\,\frac{Q^2}{k}
\ .
\ee
The coherent state is thus determined by the occupation numbers
\be
g_M(k)
=
\sqrt{\frac{k}{2}}\,\frac{\tilde V_M}{\lp}
=
-\frac{4\,\pi\,M}{\sqrt{2\,k^3}\,\mpl}
\label{gkM}
\ee
and
\be
g_Q(k)
=
\sqrt{\frac{k}{2}}\,\frac{\tilde V_Q}{\lp}
=
\frac{\pi^2\,Q^2}{\sqrt{2\,k}\,\mpl}
\ ,
\label{gkQ}
\ee
and it finally reads
\be
\ket{g}
=
e^{-N_{\rm G}/2}\,
\exp\left\{
\int\limits_0^\infty
\frac{k^2\,\dd k}{2\,\pi^2}
\left[
g_M(k)
+
g_Q(k)
\right]
\ac_k
\right\}
\ket{0}
\ ,
\label{gstate}
\ee
where the total occupation number is given by
\be
\label{NGN}
\Ng
=
\int\limits_0^{\infty} 
\frac{k^2\,\dd k}{2\,\pi^2}
\left[
g_M(k)
+
g_Q(k)
\right]^2
\equiv
N_M+N_Q+N_{MQ}
\ .
\ee
In particular, the contribution $N_M$ associated with the ADM mass $M$ is the
same as the one for the Schwarzschild metric~\cite{Casadio:2021eio} and diverges for
the exact occupation numbers $g_M$ given in Eq.~\eqref{gkM}.
This divergence implies that the $g_M=g_M(k)$ which are realised in Nature must have
different IR and UV behaviours, and that the corresponding metric must therefore differ
from the classical expression.
\par
Assuming that we do not know the actual quantum states which are realised in Nature, we can
formally express the requirement that the state $\ket{g}$ is well defined by introducing (sharp) 
IR and UV cut-offs $k_{\rm IR}\sim R_\infty^{-1}$ and $k_{\rm UV}\sim R_{\rm UV}^{-1}$,
with $R_{\rm UV}<R_+\ll R_\infty$.
It is then important to remark that the specific functional dependences on $R_\infty$ and $R_{\rm UV}$
displayed in the following are consequences of the choice of sharp cut-offs in the momentum integrals
and should not be taken too literally. 
With this proviso, one finds
\be
N_M
=
\frac{4 \,M^2}{\mpl^2}
\int\limits_{k_{\rm IR}}^{k_{\rm UV}}
\frac{\dd k}{k}
=
\frac{4\,M^2}{\mpl^2}\,
\ln\left(\frac{R_\infty}{R_{\rm UV}}\right)
\ .
\ee
Likewise, we have
\be
N_Q = 
\frac{\pi ^2\, Q^4}{4\, \mpl^2}
\int\limits_{k_{\rm IR}}^{k_{\rm UV}}
k\,{\dd k}
=
\frac{\pi ^2\, Q^4}{8\, \mpl^2}
\left(
\frac{1}{R_{\rm UV}^2}
-\frac{1}{R_\infty^2}
\right)
\ee
and the cross term
\be
N_{MQ}
=
-\frac{2\, \pi \, M\,Q^2}{\mpl^2}
\int\limits_{k_{\rm IR}}^{k_{\rm UV}}
{\dd k}
=
-\frac{2 \,\pi \, M\,Q^2}{\mpl^2}
\left(
\frac{1}{R_{\rm UV}}
-\frac{1}{R_\infty}
\right)
\ .
\ee
\par
Another quantity of interest is the average radial momentum
\be
\label{EN}
\expec{k}
=
\int\limits_0^{\infty} 
\frac{k^2\,\dd k}{2\,\pi ^2} 
\,k
\left[
g_M(k)
+
g_Q(k)
\right]^2
\equiv
\expec{k}_M
+\expec{k}_Q
+\expec{k}_{MQ}
\ ,
\ee
where the mass contribution is given by
\be
\expec{k}_M
=
\frac{4\,M^2}{\mpl^2}
\int\limits_{k_{\rm IR}}^{k_{\rm UV}}
\dd k
=
\frac{4\,M^2}{\mpl^2}
\left(
\frac{1}{R_{\rm UV}}
-\frac{1}{R_\infty}
\right)
\ ,
\ee
the charge contribution by
\be
\expec{k}_Q
=
\frac{\pi^2\, Q^4}{4\,\mpl^2}
\int\limits_{k_{\rm IR}}^{k_{\rm UV}}
k^2\,{\dd k}
=
\frac{\pi^2 \,Q^4}{12\,\mpl^2}
\left(
\frac{1}{R_{\rm UV}^3}
-\frac{1}{R_\infty^3}
\right)
\ee
and the cross term by
\be
\expec{k}_{MQ}
=
-\frac{\pi \,M\,Q^2}{\mpl^2}
\int\limits_{k_{\rm IR}}^{k_{\rm UV}}
k\,{\dd k}
=
-\frac{\pi \, M\,Q^2}{\mpl^2}
\left(
\frac{1}{R_{\rm UV}^2}
-\frac{1}{R_\infty^2}
\right)
\ .
\ee
From the above results, one can obtain the ``typical'' wavelength $\lambda_{\rm G}$.
Specifically, assuming $R_\infty \gg R_{\rm UV}$, one finds at leading order
\be
N_{\rm G}
 \simeq
\frac{4M^2}{\mpl^2}\,
\ln\left(\frac{R_\infty}{R_{\rm UV}}\right)
+
\frac{\pi ^2\, Q^4}{8\, \mpl^2\, R_{\rm UV}^2}
-\frac{2\, \pi \, M\,Q^2}{\mpl^2 \,R_{\rm UV}}
\ee
and
\be
\expec{k} 
\simeq
\frac{4\,M^2}{\mpl^2 \,R_{\rm UV}}
+
\frac{\pi^2 \,Q^4}{12\,\mpl^2\,R_{\rm UV}^3}
-\frac{\pi \, M\,Q^2}{\mpl^2\, R_{\rm UV}^2}
\ .
\ee
If we further assume $2\,\gn\, M \gg \gn\, Q^2$, we find the approximate expression
\be
\lambda_{\rm G}
=
\frac{N_{\rm G}}{\expec{k}}
\simeq 
R_{\rm UV}
\left(
1
+
\frac{\pi\, Q^2}{4\, M\,R_{\rm UV}}
\right)
\ln\left(\frac{R_\infty}{R_{\rm UV}}\right) 
\ ,
\ee
which reduces to the Schwarzschild case~\cite{Casadio:2021eio} for $Q=0$.
The effect of (a relatively small) charge is therefore to increase the typical
wavelength of the gravitons building the geometry.
\section{Quantum corrected Reissner-Nordstr\"om spacetime}
\label{s:qmetric}
\begin{figure}[t]
\centering
\includegraphics[width=8cm]{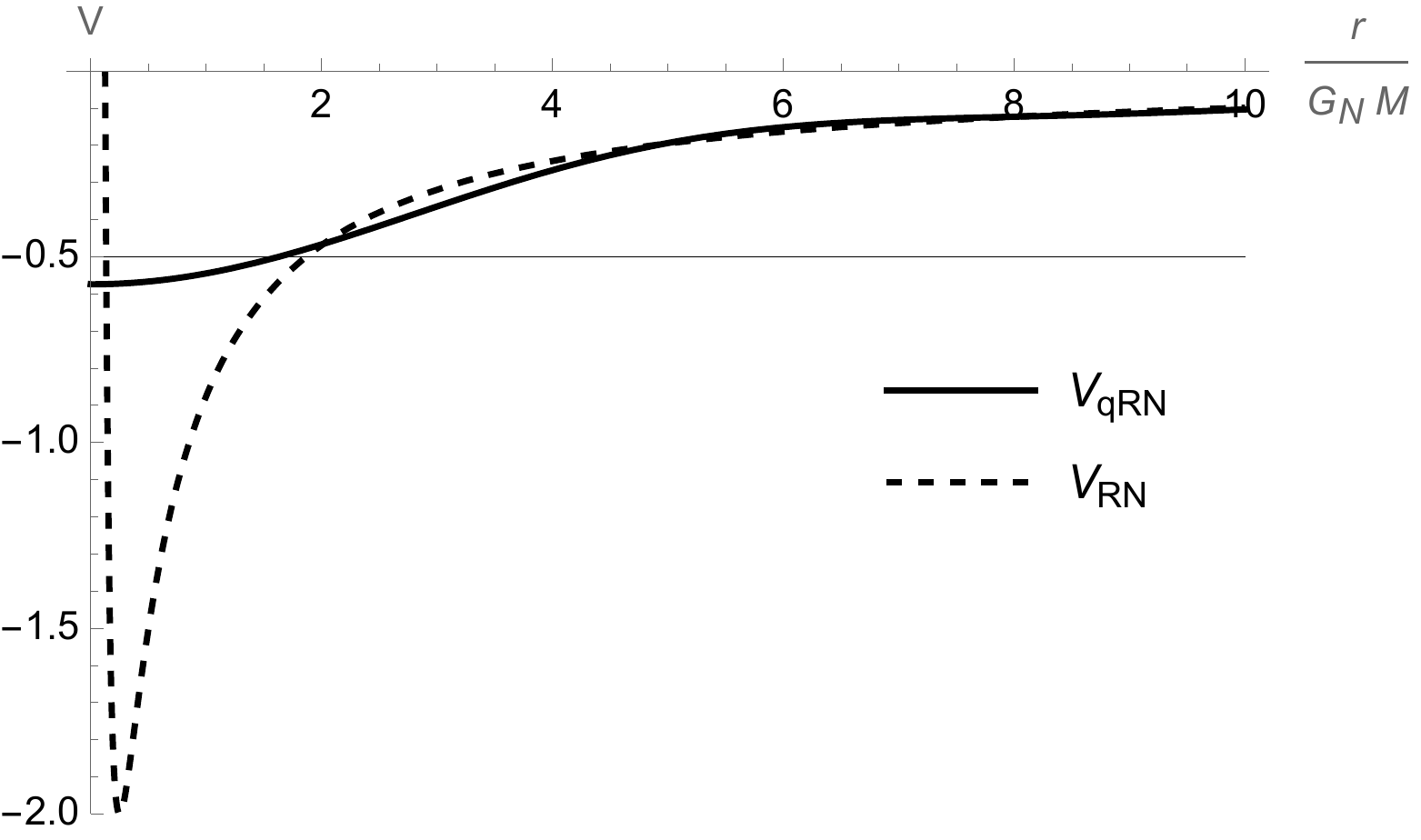}
$\ $
\includegraphics[width=8cm]{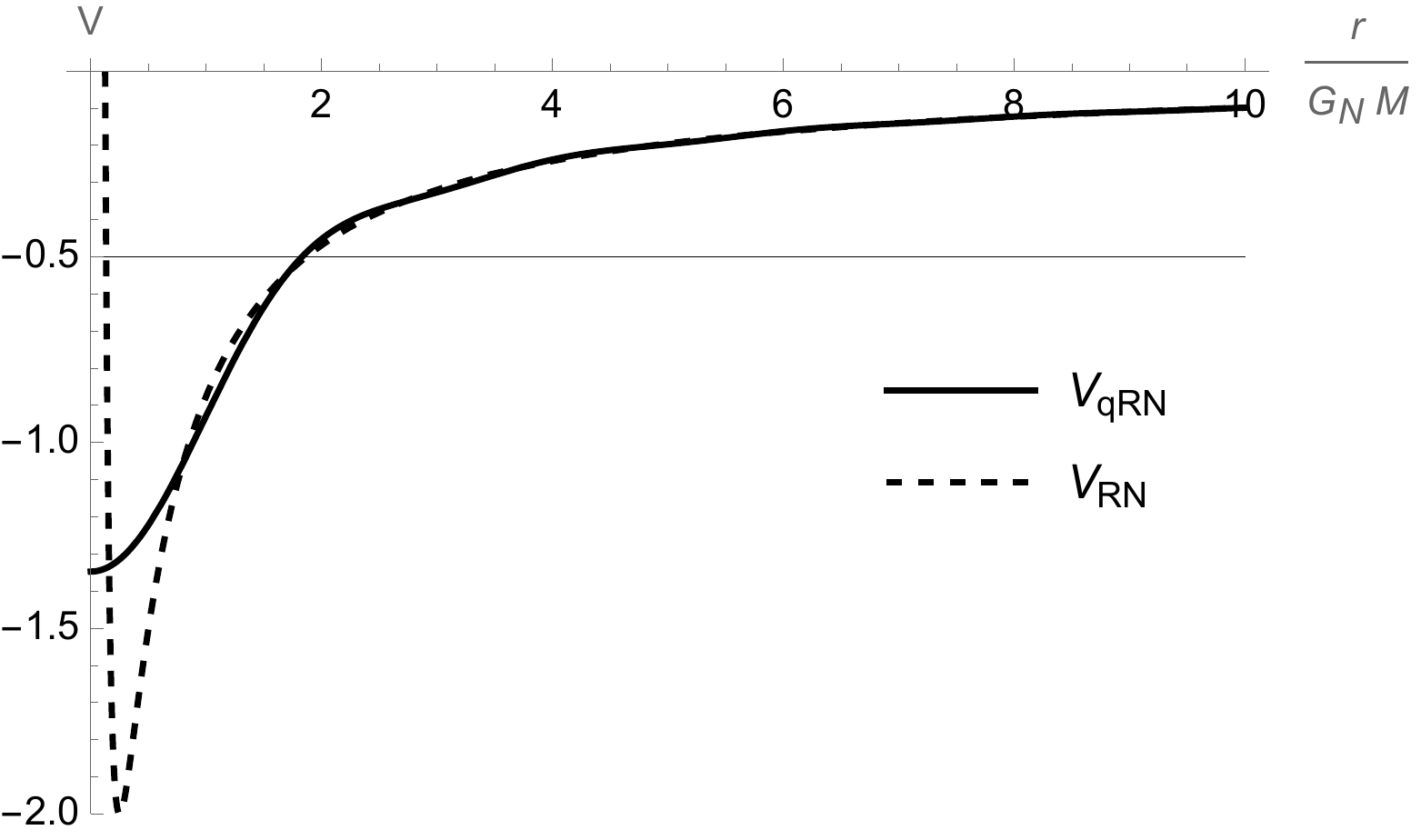}
\caption{Quantum potential $V_{\rm qRN}$ in Eq.~\eqref{VqRN} (solid line) compared to $V_{\rm RN}$
(dashed line) for $R_{\rm UV}=\gn\,M=2\,\sqrt{\gn\,Q^2}$ (left panel) and
for $R_{\rm UV}=\gn\,M/3=2\,\sqrt{\gn\,Q^2}/3$ (right panel).
The thin solid line $V=-1/2$ crosses the potential at the horizons.}
\label{pVqRN}
\end{figure}

\begin{figure}[h]
\centering
\includegraphics[width=8cm]{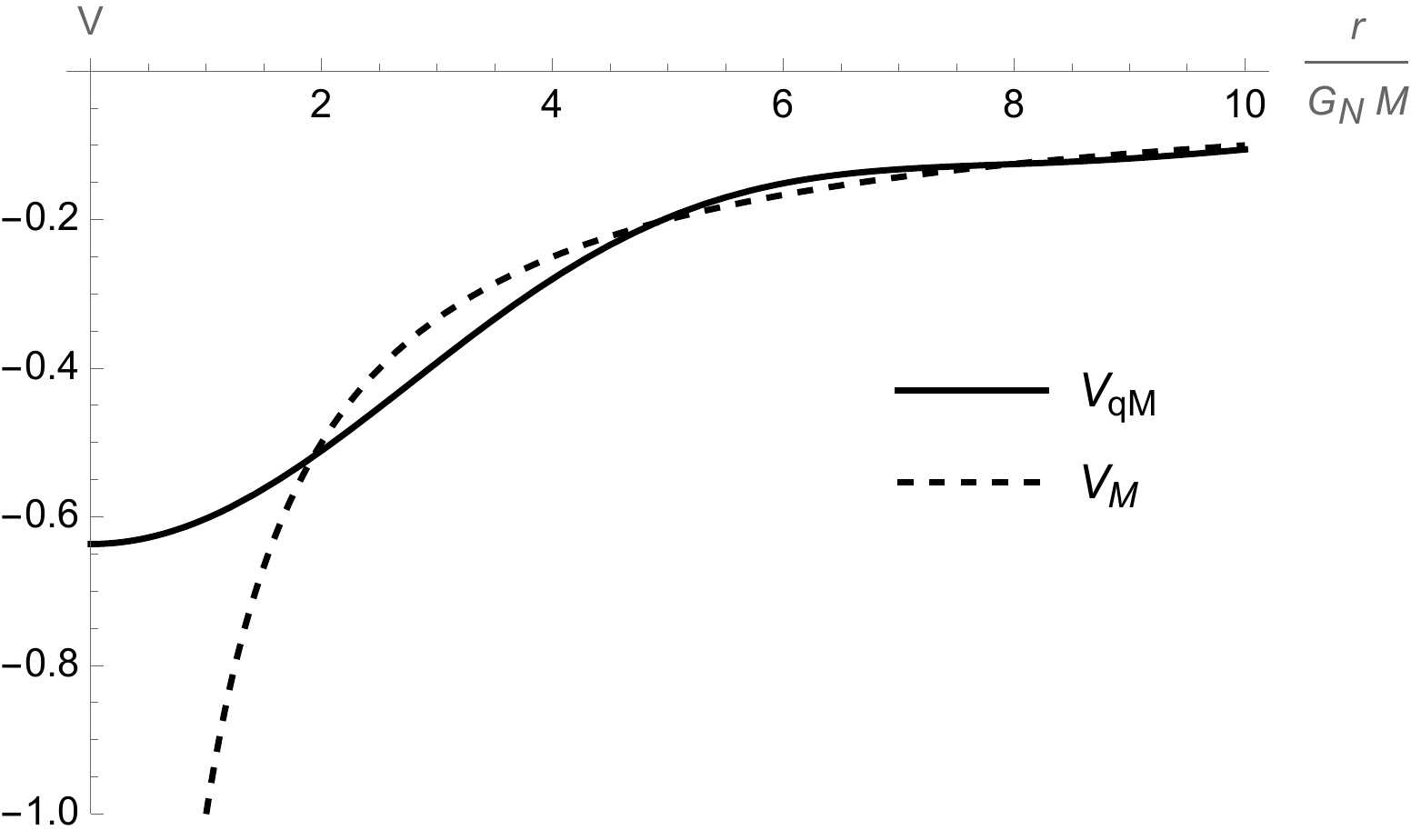}
$\ $
\includegraphics[width=8cm]{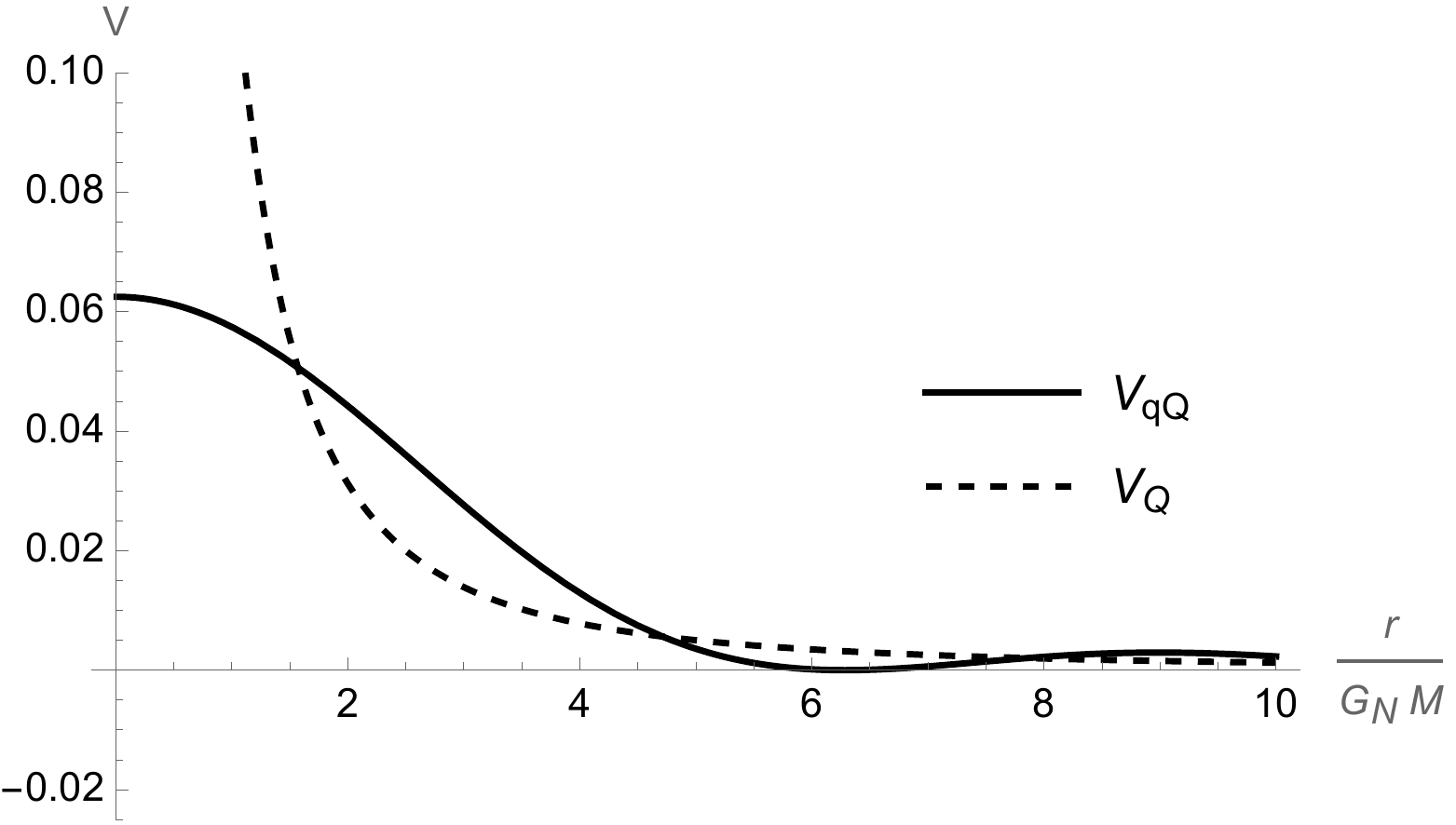}
\caption{Quantum potential $V_{{\rm q}M}$ in Eq.~\eqref{VqM} (solid line) compared to $V_{M}$
(dashed line) for $R_{\rm UV}=\gn\,M=2\,\sqrt{\gn\,Q^2}$ (left panel) and
$V_{{\rm q}Q}$ in Eq.~\eqref{VqQ} (solid line) compared to $V_{Q}$
for $R_{\rm UV}=\gn\,M/3=2\,\sqrt{\gn\,Q^2}/3$ (right panel).}
\label{pVVqRN}
\end{figure}

\setcounter{equation}{0}
We can now reconstruct the quantum corrected metric functions by simply integrating
back the occupation numbers~\eqref{gkM} and \eqref{gkQ} between the IR and UV cut-offs,
which yields
\be
V_{{\rm q}M}
=
-\frac{2\,\gn\,M}{\pi\,r}
\left[
\Si\!\left(\frac{r}{R_{\rm UV}}\right)
-\Si\!\left(\frac{r}{R_\infty}\right)
\right]
\ ,
\label{VqM}
\ee
where $\Si=\Si(x)$ denotes the sine integral function, and
\be
V_{{\rm q}Q}
=
\frac{\gn\,Q^2}{2\,r^2}
\left[
\cos\!\left(\frac{r}{R_\infty}\right)
-\cos\!\left(\frac{r}{R_{\rm UV}}\right)
\right]
\ .
\label{VqQ}
\ee
In both expressions, we can safely take the limit $R_\infty\to\infty$ as an approximation,
so that we finally obtain
\be
V_{{\rm qRN}}
\simeq
-\frac{2 \,\gn\,M}{\pi\,r}\,
\Si\!\left(\frac{r}{R_{\rm UV}}\right)
+
\frac{\gn\,Q^2}{2\,r^2}
\left[
1
-\cos\!\left(\frac{r}{R_{\rm UV}}\right)
\right]
\ .
\label{VqRN}
\ee
Two examples of the corrected potential are plotted in Fig.~\ref{pVqRN}.
We can in particular notice that the oscillations around the classical mass contribution
asymptote to zero,~\footnote{We recall that $\lim_{x\to\infty} {\rm Si}(x)=\pi/2$.}
whereas the oscillations around the term containing the charge $Q$ have
constant amplitude (see Fig.~\ref{pVVqRN}). 
\par
We next use the function~\eqref{VqRN} to define the quantum corrected metric
\be
\d s^2
\simeq
-\left(1+2\,V_{\rm qRN}\right)\d t^2
+
\frac{\d r^2}{1+2\,V_{\rm qRN}}
+
r^2\,\d\Omega^2
\ ,
\label{gqRN}
\ee
where the approximate equality is to remind us of all the simplifying assumptions, including
the fact that we have neglected the IR departure from the classical behaviour.
Strictly speaking, the above approximation is valid as long as we consider $R_{\rm UV}\ll r\ll R_\infty$.
Nonetheless, we will investigate the entire region $r>0$ in the following.
\subsection{Effective source}
\begin{figure}[t]
\centering
\includegraphics[width=8cm]{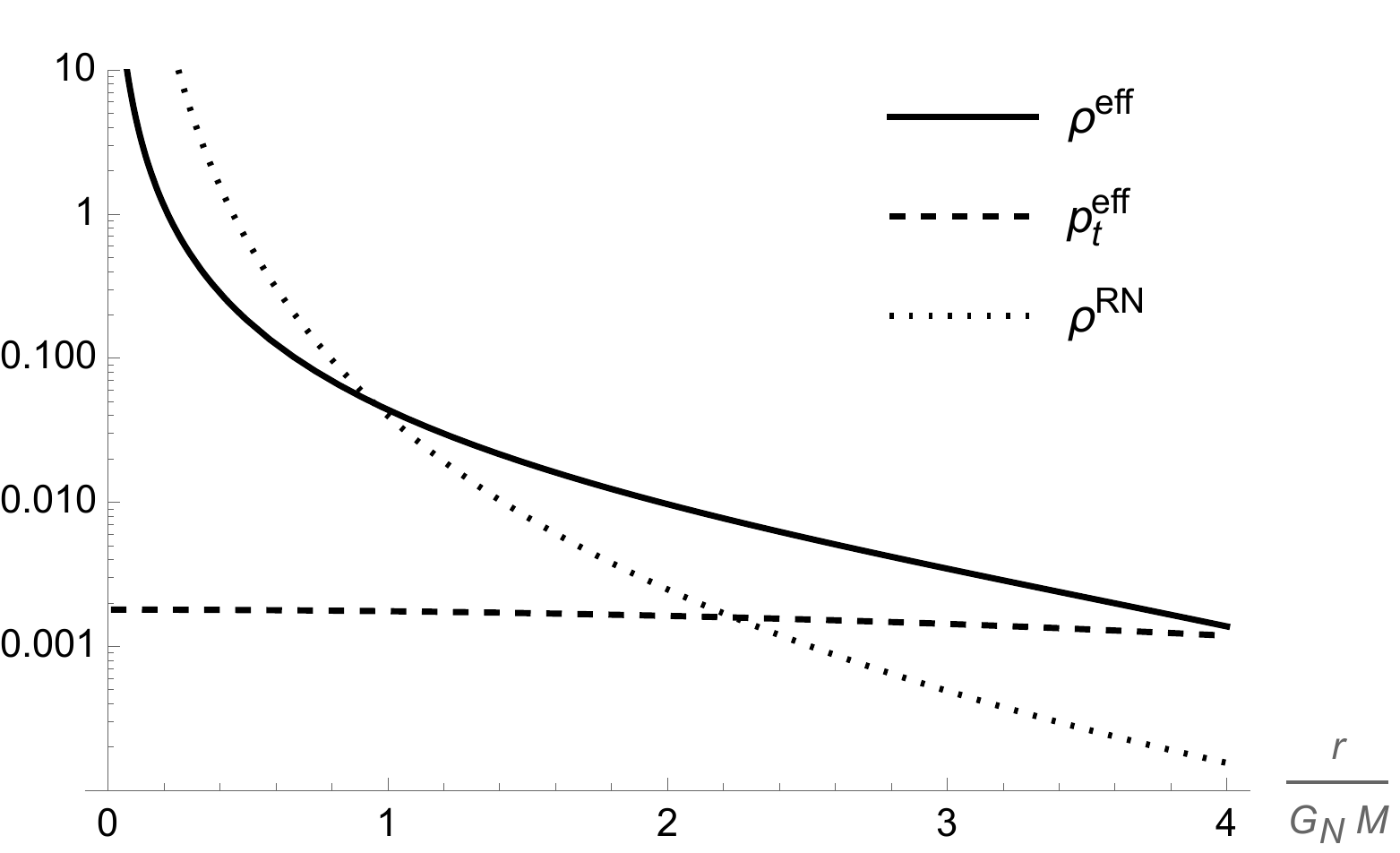}
$\ $
\includegraphics[width=8cm]{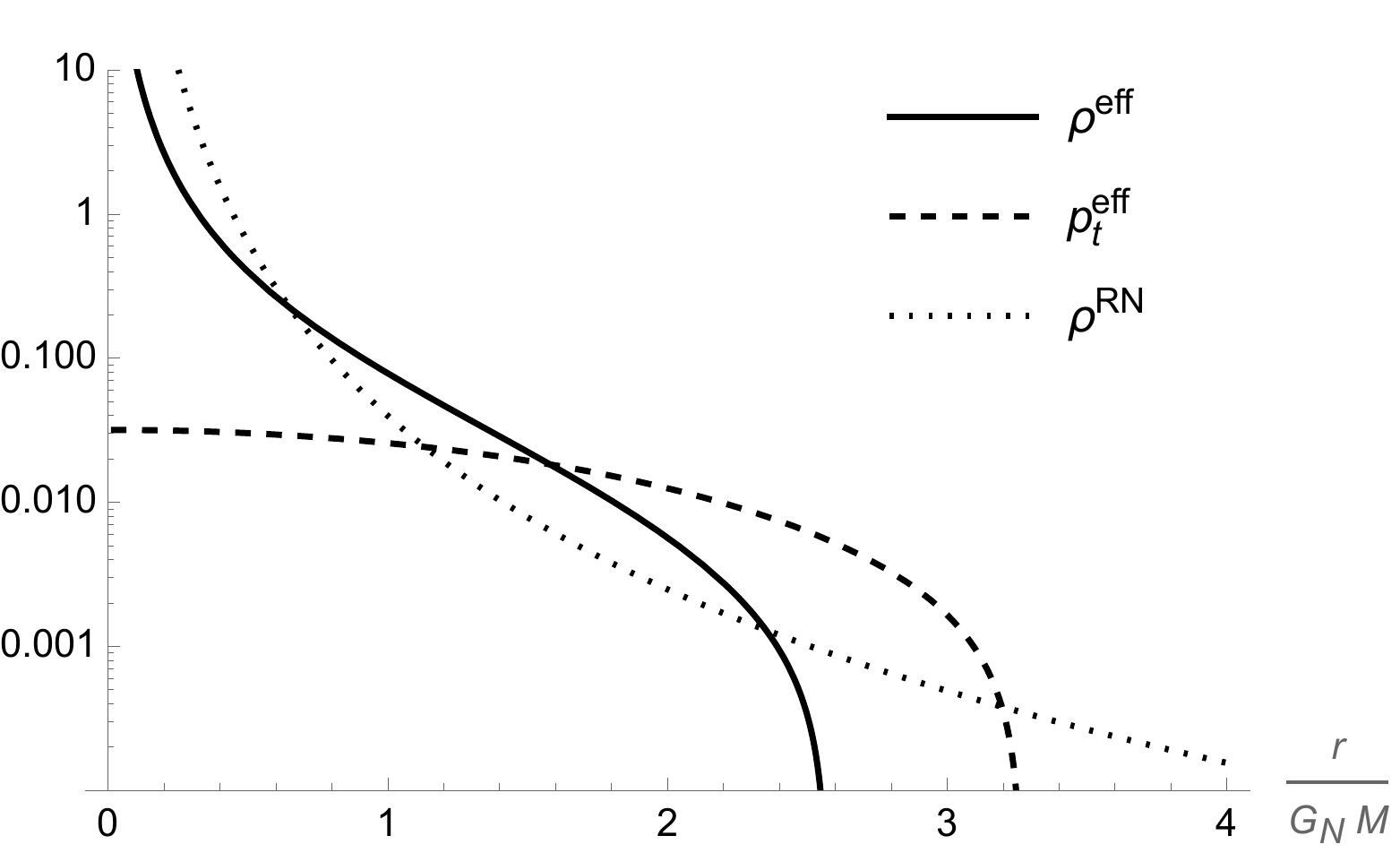}
\caption{Effective energy density (solid line) and tangential pressure (dashed line) in Eq.~\eqref{Teff} 
compared to the Reissner-Nordstr\"om contribution~\eqref{Trn} (dotted line)
for $R_{\rm UV}=\gn\,M=2\,\sqrt{\gn\,Q^2}$ (left panel) and
for $R_{\rm UV}=\gn\,M/3=2\,\sqrt{\gn\,Q^2}/3$ (right panel).}
\label{pTqRN}
\end{figure}

\begin{figure}[t]
\centering
\includegraphics[width=8cm]{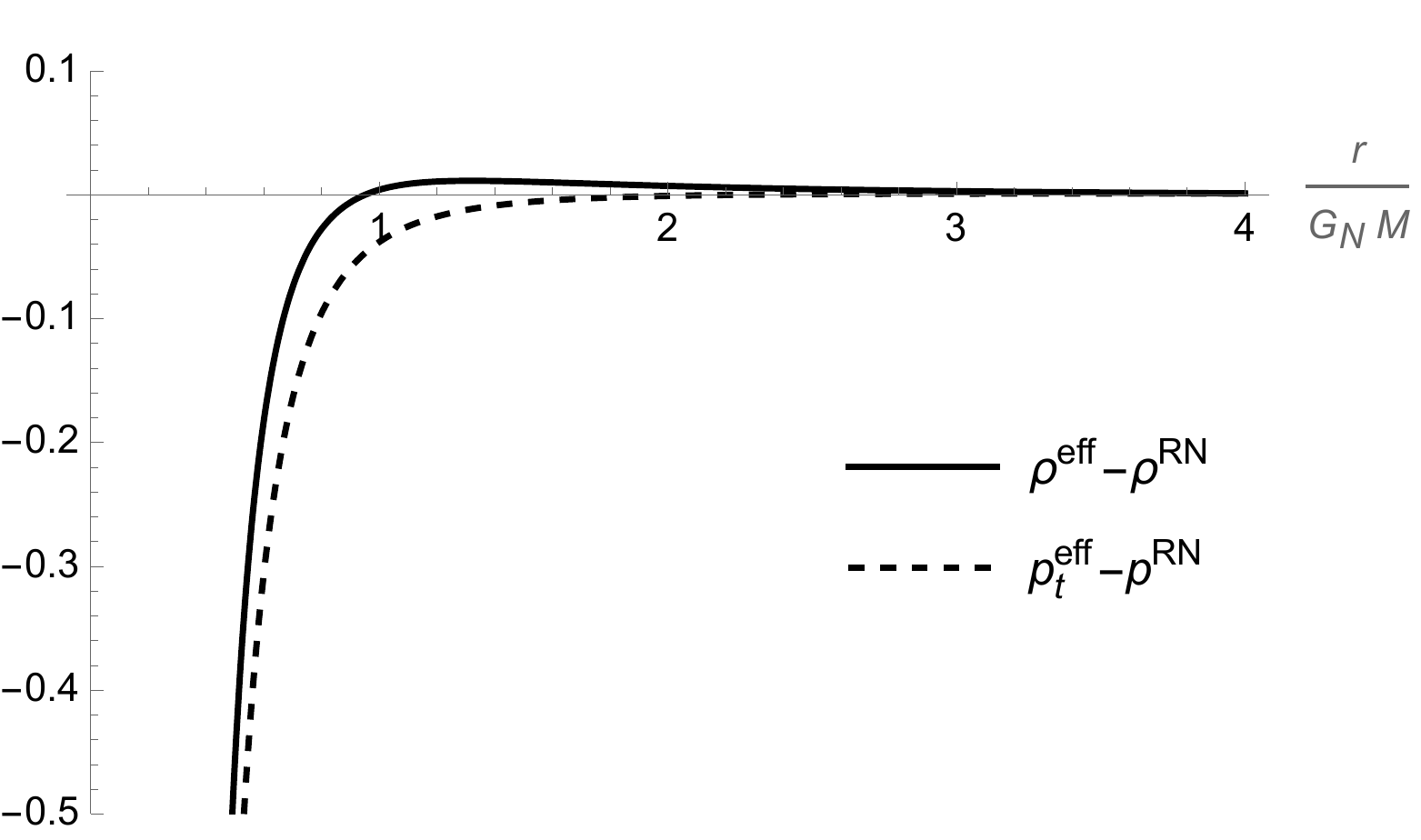}
$\ $
\includegraphics[width=8cm]{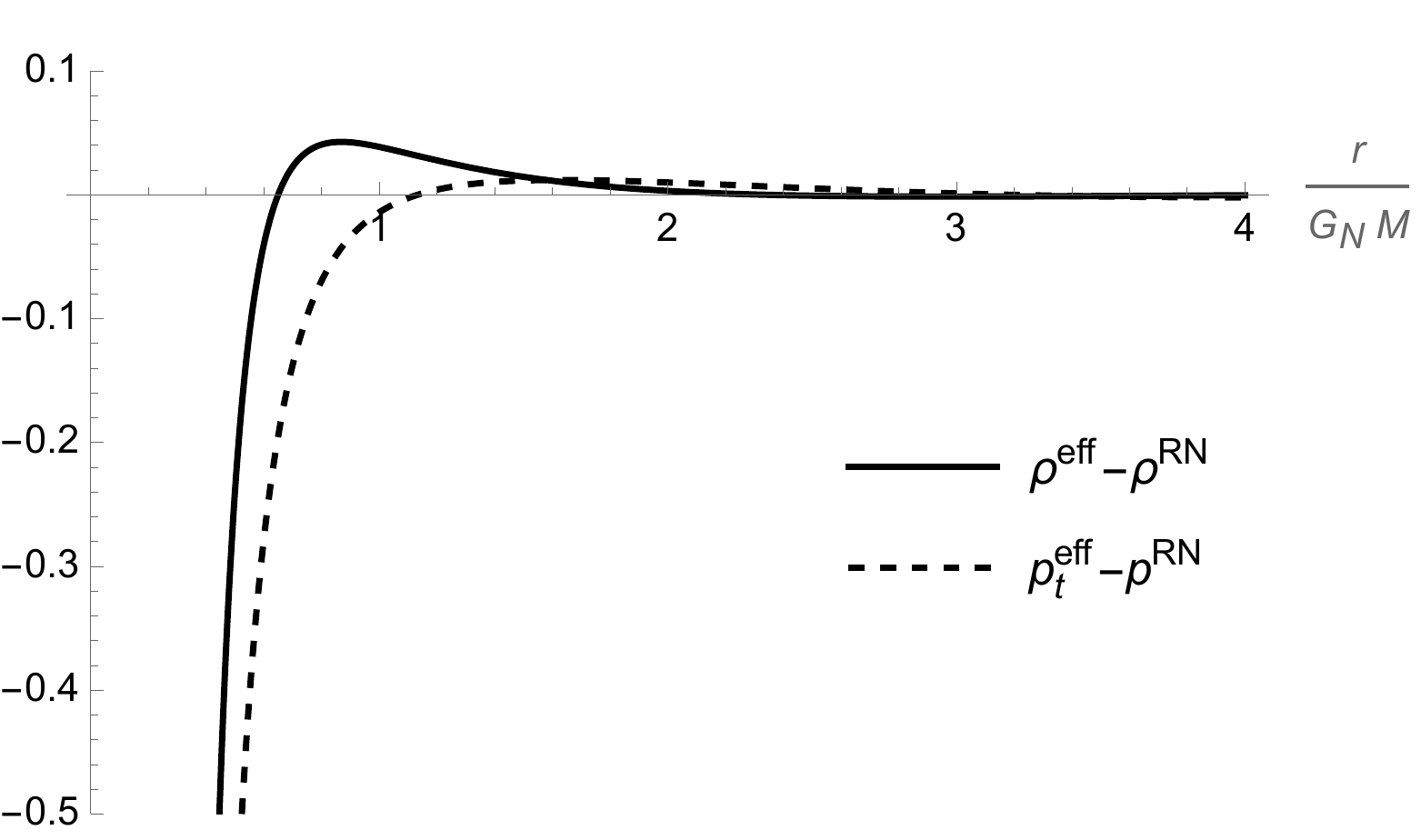}
\caption{Quantum contribution to the effective energy density (solid line) and tangential pressure (dashed line) in Eq.~\eqref{Teff} 
for $R_{\rm UV}=\gn\,M=2\,\sqrt{\gn\,Q^2}$ (left panel) and
for $R_{\rm UV}=\gn\,M/3=2\,\sqrt{\gn\,Q^2}/3$ (right panel).}
\label{pTTqRN}
\end{figure}

From the point of view of general relativity, both the classical Reissner-Nordstr\"om 
metric~\eqref{gRN} and Eq.~\eqref{gqRN} are not solutions in the vacuum.
The (effective) Einstein equations are sourced by an (effective) energy-momentum tensor
\be
T^\mu_{\ \nu}
=
{\rm diag}
\left(
-\rho^{\rm eff},
p^{\rm eff}_r,
p^{\rm eff}_t,
p^{\rm eff}_t
\right)
=
\frac{G^\mu_{\ \nu}}{8\,\pi\,\gn}
\ .
\label{Teff}
\ee
One can then compute the Einstein tensor $G_{\mu\nu}$ for the metric~\eqref{gqRN} in order to determine
the effective energy density
\be
\label{de}
\rho^{\rm eff}
=
\frac{Q^2}{8 \,\pi\,  r^4}
\left[
1-\cos \left(\frac{r}{R_{\rm UV}}\right)
\right]
+\frac{4\,M\,R_{\rm UV}-\pi\,Q^2}{8\, \pi ^2\, r^3\,R_{\rm UV}}\,
\sin\left(\frac{r}{R_{\rm UV}}\right)
\ee
the radial pressure
\be
\label{pr}
p^{\rm eff}_r
=
-\rho^{\rm eff}
\ ,
\ee
and the tension
\be
\label{pt}
p^{\rm eff}_t
&\!\!=\!\!&
\frac{Q^2}{8\,\pi\,r^4}
\left[
1-\cos \left(\frac{r}{R_{\rm UV}}\right)
\right]
\nonumber
\\
&\!\!\!\!&
+
\frac{\left(\pi  \,Q^2
-4 \,M \,R_{\rm UV}\right)
r\,\cos \left(\frac{r}{R_{\rm UV}}\right)
+2 \,R_{\rm UV} \left(2\, M\, R_{\rm UV}-\pi\,  Q^2\right) \sin \left(\frac{r}{R_{\rm UV}}\right)}
{16 \,\pi ^2 \,r^3 \,R_{\rm UV}^2}
\ .
\ee
In the above expressions, the first term (independent of the cut-off $R_{\rm UV}$)
is the standard (traceless) electrostatic contribution for the Reissner-Nordstr\"om metric,
\be
\rho^{\rm RN}
=
-p_r^{\rm RN}
=
p_t^{\rm RN}
=
\frac{Q^2}{8\,\pi\,r^4}
\ .
\label{Trn}
\ee
The additional terms in Eq.~\eqref{de},
\be
\label{dde}
\rho^{\rm eff}
-
\rho^{\rm RN}
=
\frac{M}{2\, \pi ^2\, r^3}\,
\sin\left(\frac{r}{R_{\rm UV}}\right)
-
\frac{Q^2}{8 \,\pi\,  r^4}
\left[
\cos \left(\frac{r}{R_{\rm UV}}\right)
+
\frac{r}{R_{\rm UV}}\,
\sin\left(\frac{r}{R_{\rm UV}}\right)
\right]
\ ,
\ee
and those in Eq.~\eqref{pt},
\be
\label{dpt}
p^{\rm eff}_t
-
p_t^{\rm RN}
&\!\!=\!\!&
\frac{M}{4 \,\pi ^2 \,r^3}
\left[
 \sin \left(\frac{r}{R_{\rm UV}}\right)
-
\frac{r}{R_{\rm UV}}\,
\cos \left(\frac{r}{R_{\rm UV}}\right)
\right]
\nonumber
\\
&\!\!\!\!&
-
\frac{Q^2}{8\,\pi\,r^4}
\left[
\left(1
-
\frac{r^2}{2\,R_{\rm UV}^2}
\right)
\cos \left(\frac{r}{R_{\rm UV}}\right)
+
\frac{r}{R_{\rm UV}}\,
\sin \left(\frac{r}{R_{\rm UV}}\right)
\right]
\ ,
\ee
can therefore be interpreted as representing an effective (quantum) smearing of both the mass
and the charge of the central source over the length scale $R_{\rm UV}$.
The energy density and pressures corresponding to the cases in Figs.~\ref{pVqRN} and
\ref{pVVqRN} are displayed in Figs.~\ref{pTqRN} and Fig.~\ref{pTTqRN}.
\par
\begin{figure}[t]
\centering
\includegraphics[width=8cm]{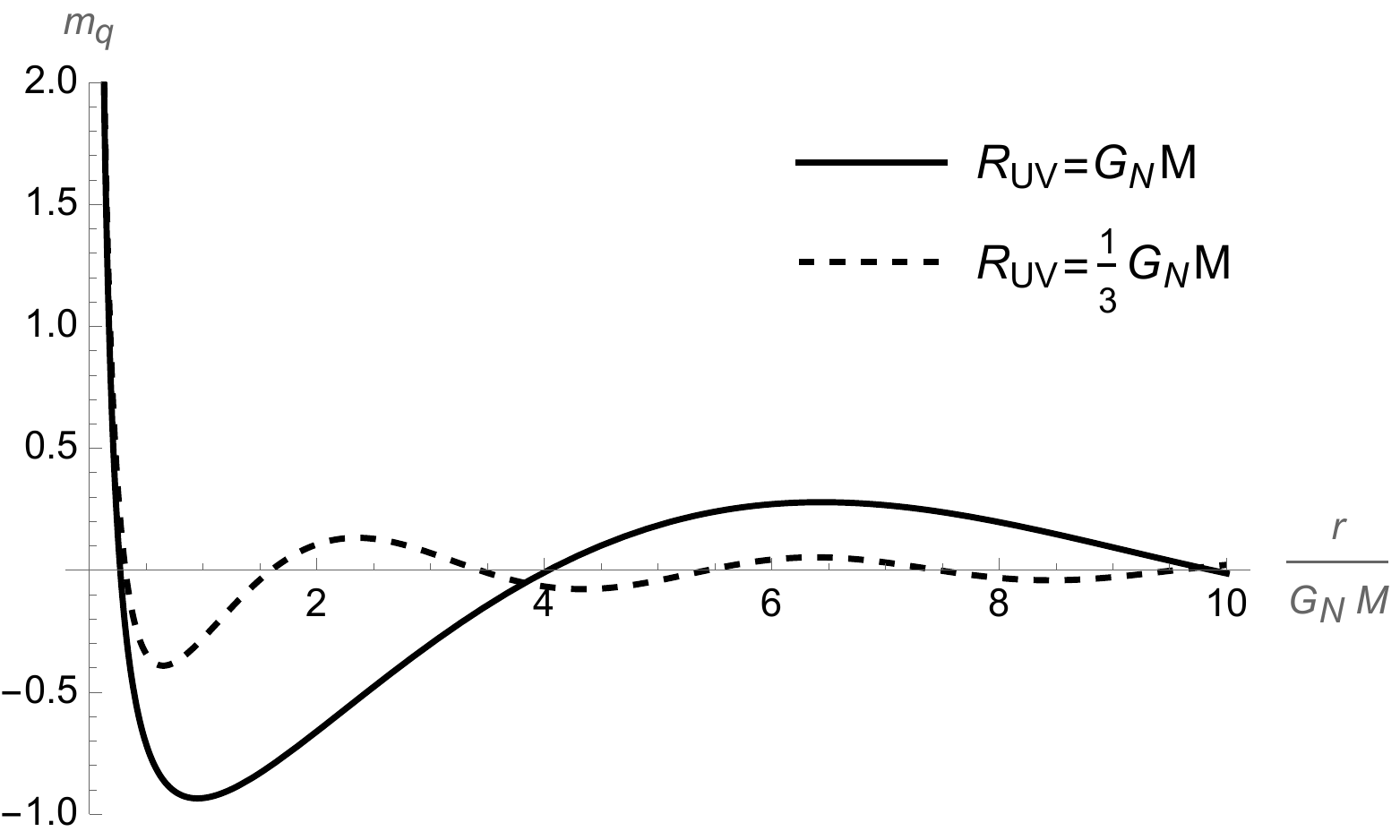}
\caption{Mass function in Eq.~\eqref{Mq} 
for $R_{\rm UV}=\gn\,M=2\,\sqrt{\gn\,Q^2}$ (solid linel) and
for $R_{\rm UV}=\gn\,M/3=2\,\sqrt{\gn\,Q^2}/3$ (dashed line).}
\label{pMq}
\end{figure}

The energy-momentum tensor~\eqref{Teff} satisfies the the conservation equation
$\nabla_\mu T^{\mu\nu} = 0$ by construction.
In particular, the only non-trivial condition is for $\nu=1$ and yields the Tolman-Oppenheimer-Volkoff (TOV)
equation
\be
\label{con111}
\frac{\d p^{\rm eff}_r}{\d r}
=
-\frac{1}{1+2\,V_{\rm qRN}}\frac{\d V_{\rm qRN}}{\d r}\,
\left(\rho^{\rm eff}+p^{\rm eff}_r\right)
+
\frac{2}{r}\left(p^{\rm eff}_t-p^{\rm eff}_r\right)
\ ,
\ee
which governs the hydrodynamic equilibrium of the system. 
In particular, the quantum corrected metric~\eqref{gqRN} is still of the Kerr-Schild form~\cite{KS}, 
and the effective fluid is anisotropic with $p^{\rm eff}_t\not=p^{\rm eff}_r$.
We can then note that the metric~\eqref{gqRN} can be formally thought as
the coupling~\cite{Ovalle:2017fgl,Ovalle:2018gic} of two mass functions in a Kerr-Schild spacetime, 
namely
\be
\tilde{m}
=
m_{\rm RN}
+
m_{\rm q}
\ ,
\ee
where the mass function of the Reissner-Nordstr\"om solution is given by
\be
m_{\rm RN}
=
M-\frac{Q^2}{2\,r}
\ ,
\ee
while
\be
m_{\rm q}
=
M\left[\frac{2}{\pi}\Si\!\left(\frac{r}{R_{\rm UV}}\right)
-1
\right]
+
\frac{Q^2}{2\,r}\cos\!\left(\frac{r}{R_{\rm UV}}\right) 
\label{Mq}
\ee
is the mass function of the effective quantum fluid filling the spacetime
(see Fig.~\ref{pMq}).
\par
We next proceed to study the causal structure of the quantum corrected metric.
\subsection{Singularity}
\label{ss:sin}
We can start by analysing the metric~\eqref{gqRN} near $r=0$.
In particular, we find
\be
V_{\rm qRN}(0)
=
\frac{\gn\left(\pi\,Q^2-8\,M\,R_{\rm UV}\right)}{4\,\pi\,R_{\rm UV}^2}
\label{VqRN0}
\ee
and
\be
\left.\frac{\d V_{\rm qRN}}{\d r}\right|_{r=0}
=
0
\ ,
\ee
from which we expect no central singularity. 
In fact, we can compute the Ricci scalar to leading order around $r=0$,
\be
R
\simeq
\frac{\gn\left(8\,M\,R_{\rm UV}-\pi\,Q^2\right)}{\pi\,R_{\rm UV}^2\,r^2}
\sim
-\frac{V_{\rm qRN}(0)}{r^2}
\ee
and the Kretschmann scalar
\be
R_{\alpha\beta\mu\nu}\,R^{\alpha\beta\mu\nu}\simeq R^2\sim r^{-4}
\ .
\ee
The centre of the system is therefore an integrable singularity~\cite{lukash}
where tidal forces remain finite and the volume integral of the Ricci
scalar is also finite.
This is to be compared with the standard results
$R^2\sim R_{\alpha\beta\mu\nu}\,R^{\alpha\beta\mu\nu}\sim Q^4/r^8$
for the Reissner-Nordstr\"om metric~\eqref{gRN}.
\par
The fact that the origin is regular is further supported by the 
expressions of the energy density for $r\sim 0$, that is
\be
\label{vv1}
\rho^{\rm eff}
=
-p^{\rm eff}_r
\simeq
\frac{8\,M\,R_{\rm UV}-\pi\,Q^2}{16\,\pi^2\,r^2\,R_{\rm UV}^2}
\ ,
\ee
and the analogue expression for the tension
\be
p_t^{\rm eff}
\label{vv2}
\simeq
\frac{16\,M\,R_{\rm UV}-3\,\pi\,Q^2}{192\,\pi^2\,R_{\rm UV}^4}
\  .
\ee
Many regular black holes violate some energy conditions for $r\sim0$~\cite{regular}.
However, from Eqs.~\eqref{vv1} and~\eqref{vv2}, we can see that for
\be
16\,M\,R_{\rm UV}\geq\,3\,\pi\,Q^2
\  ,
\ee
the strong energy condition is satisfied.
\subsection{Event and Cauchy horizons}
\label{ss:ch}
\begin{figure}[t]
\centering
\includegraphics[width=8cm]{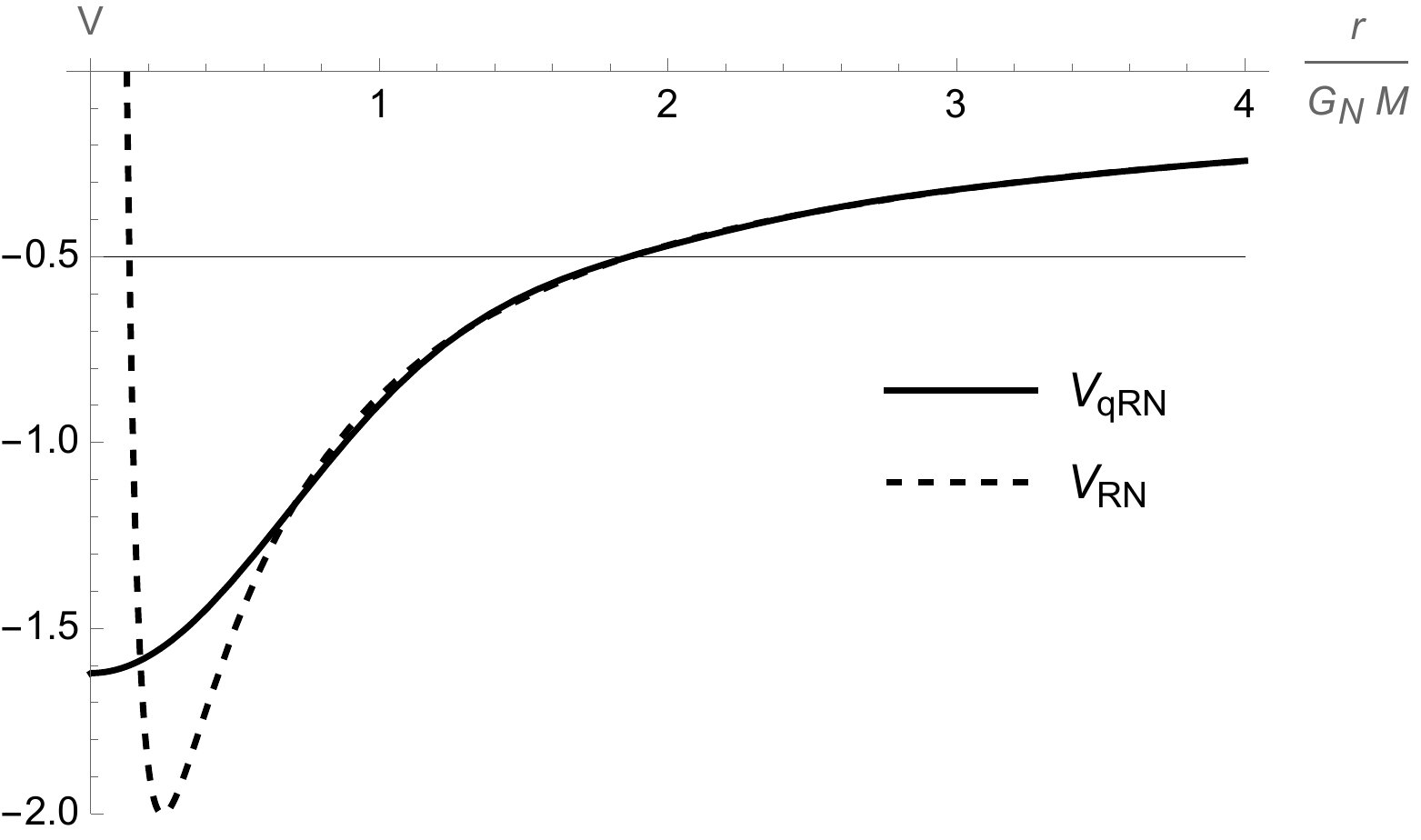}
$\ $
\includegraphics[width=8cm]{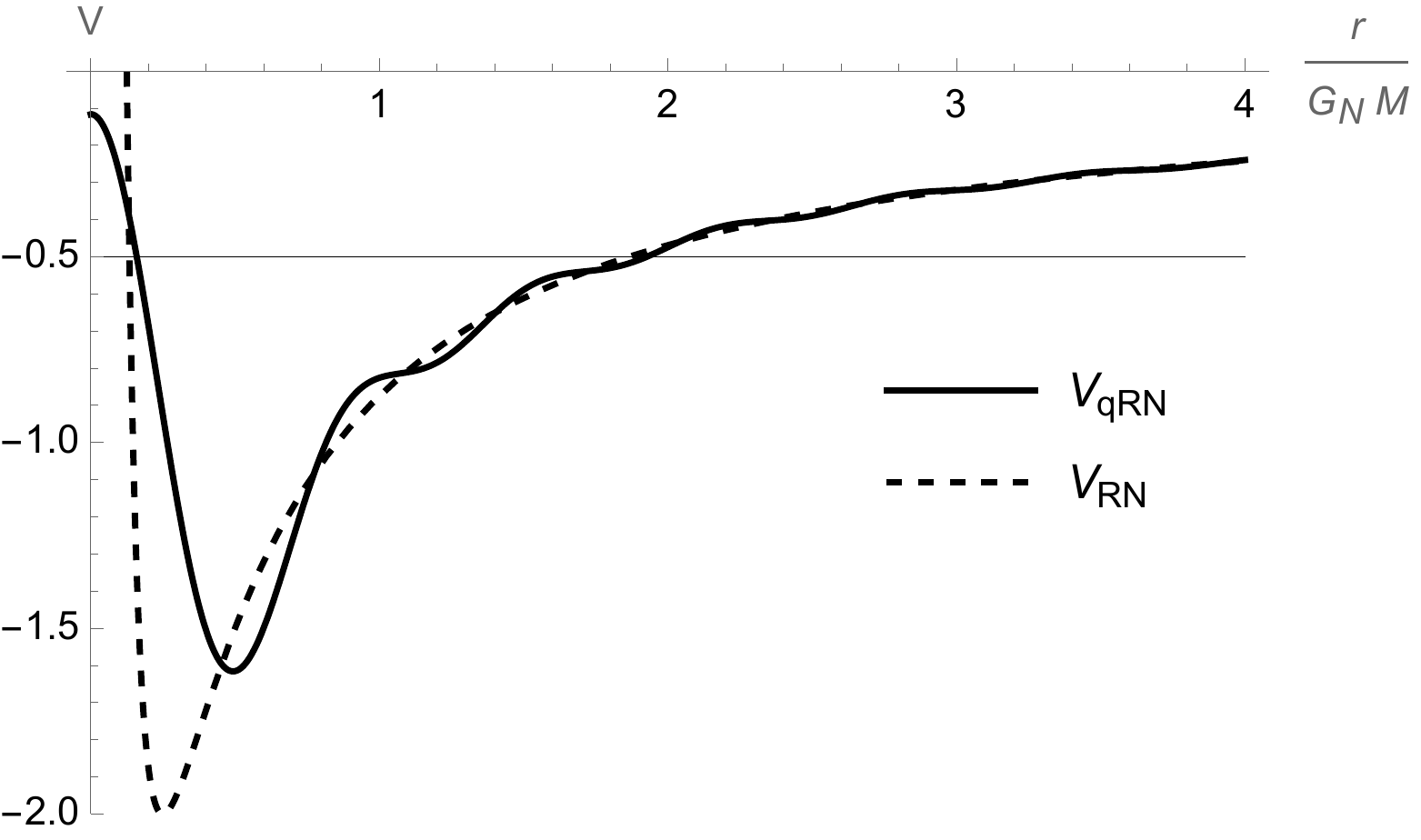}
\caption{Quantum potential $V_{\rm qRN}$ in Eq.~\eqref{VqRN} (solid line) compared to $V_{\rm RN}$
(dashed line) for $R_{\rm UV}=\gn\,M/5=2\,\sqrt{\gn\,Q^2}/5>R_-\simeq 0.13\,\gn\,M$ (left panel) and
for $R_{\rm UV}=\gn\,M/10=\sqrt{\gn\,Q^2}/5<R_-$ (right panel).
The thin solid line $V=-1/2$ crosses the potential at the horizons.}
\label{ch}
\end{figure}
\begin{figure}[t]
\centering
\includegraphics[width=8cm]{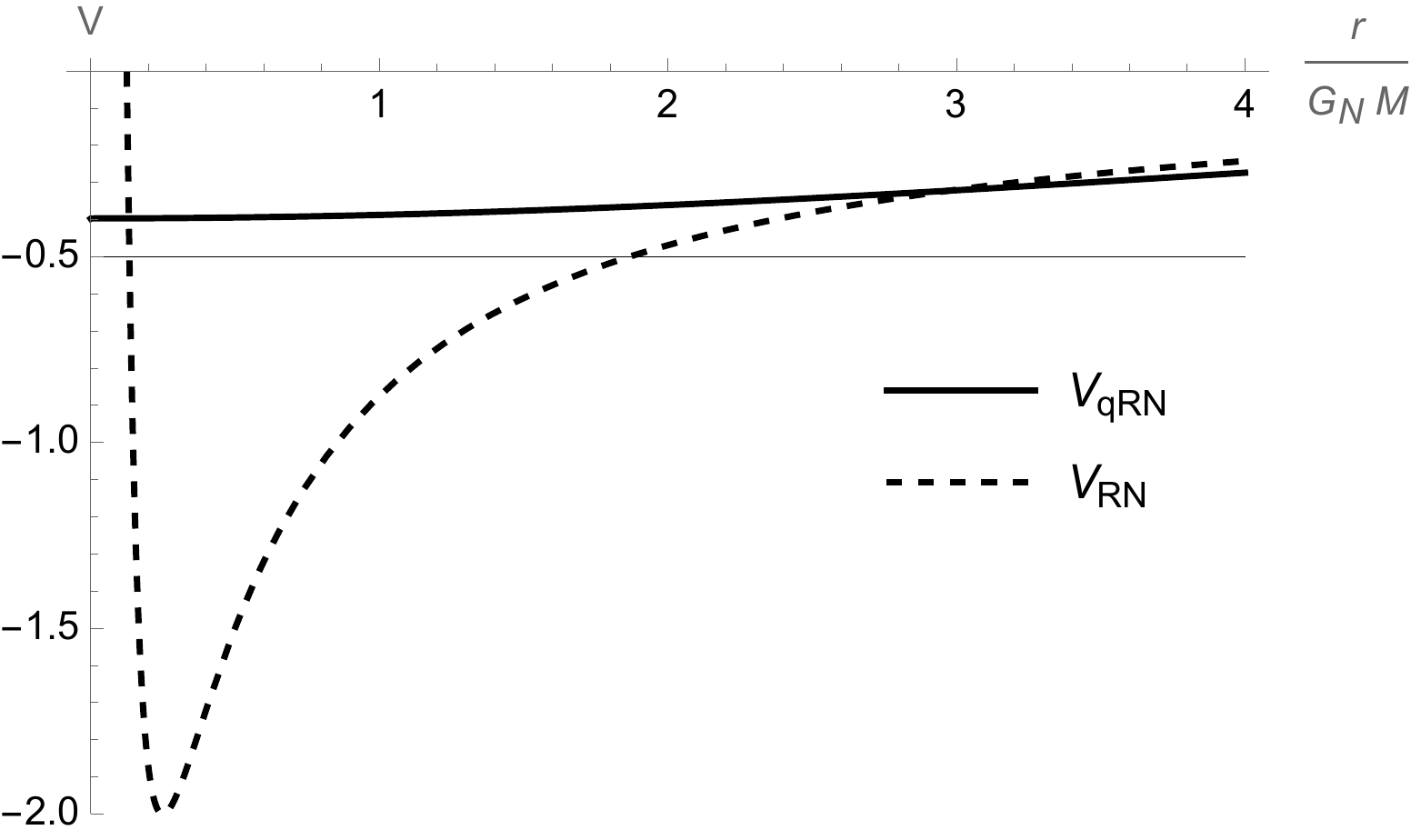}
$\ $
\includegraphics[width=8cm]{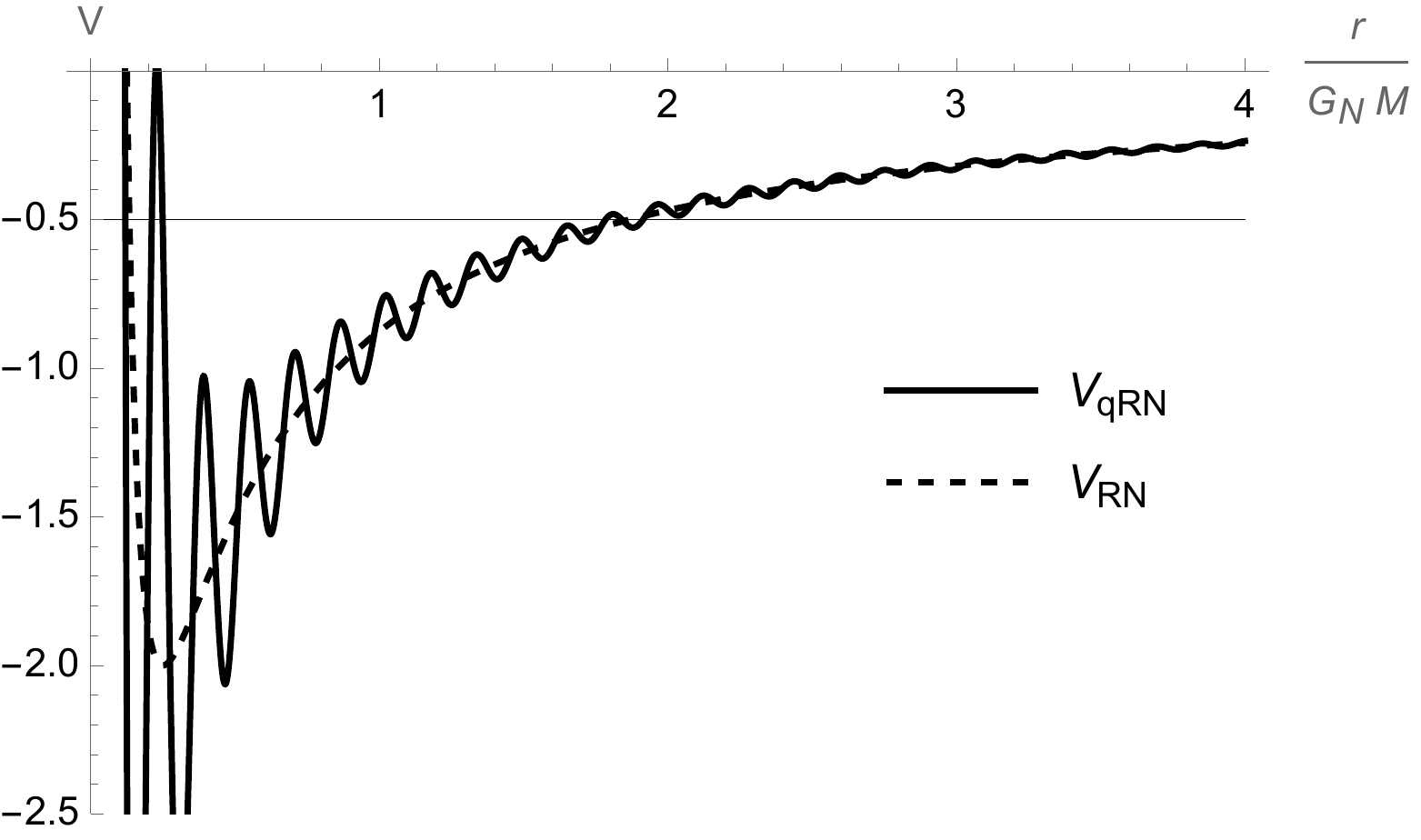}
\caption{Quantum potential $V_{\rm qRN}$ in Eq.~\eqref{VqRN} (solid line) compared to $V_{\rm RN}$
(dashed line) for $R_{\rm UV}=3\,\gn\,M/2=3\,\sqrt{\gn\,Q^2}\simeq R_+\simeq 1.8\,\gn\,M$ (left panel) and
for $R_{\rm UV}=\gn\,M/50=\sqrt{\gn\,Q^2}/25\ll R_-\simeq 0.13\,\gn\,M$ (right panel).
The thin solid line $V=-1/2$ crosses the potential at the horizons.}
\label{chX}
\end{figure}
The metric~\eqref{gqRN} can contain horizons determined by 
\be
g^{rr}
=
-g_{tt}
=
1+2\,V_{\rm qRN}=0
\ ,
\label{Rq+-}
\ee
the largest zero being the event horizon $R_{\rm q+}$ analogous to $R_+$ in Eq.~\eqref{R+-}. 
It is not possible to find analytical expressions for the above zeros for general values of
$M$, $Q$ and $R_{\rm UV}$ (see also Appendix~\ref{A:mass}).
A simple numerical inspection shows that $R_{\rm q+}$ exists in general if $R_{\rm UV}\lesssim R_+$
and $Q^2$ is such that $R_-\ll  R_+$.
We are then particularly interested in the existence of the inner Cauchy horizon, that is a second zero
$R_{\rm q-}< R_{\rm q+}$, when the event horizon $R_{\rm q+}$ also exists.
Again, a simple numerical analysis shows that no Cauchy horizon exists if $R_-\lesssim R_{\rm UV}\lesssim R_+$,
whereas there can be a inner horizon for $R_{\rm UV}\lesssim R_-$ (see left and right panels
in Fig.~\ref{ch}, respectively).
In Fig.~\ref{chX} we also show that there is no event horizon for $R_{\rm UV}\gtrsim R_+$ (left panel)
and there can be multiple inner horizons for $R_{\rm UV}\ll R_-$ (right panel).
Since there is no central singularity, the former case would represent an electrically
charged star.
\par
\begin{figure}[t]
\centering
\includegraphics[width=8cm]{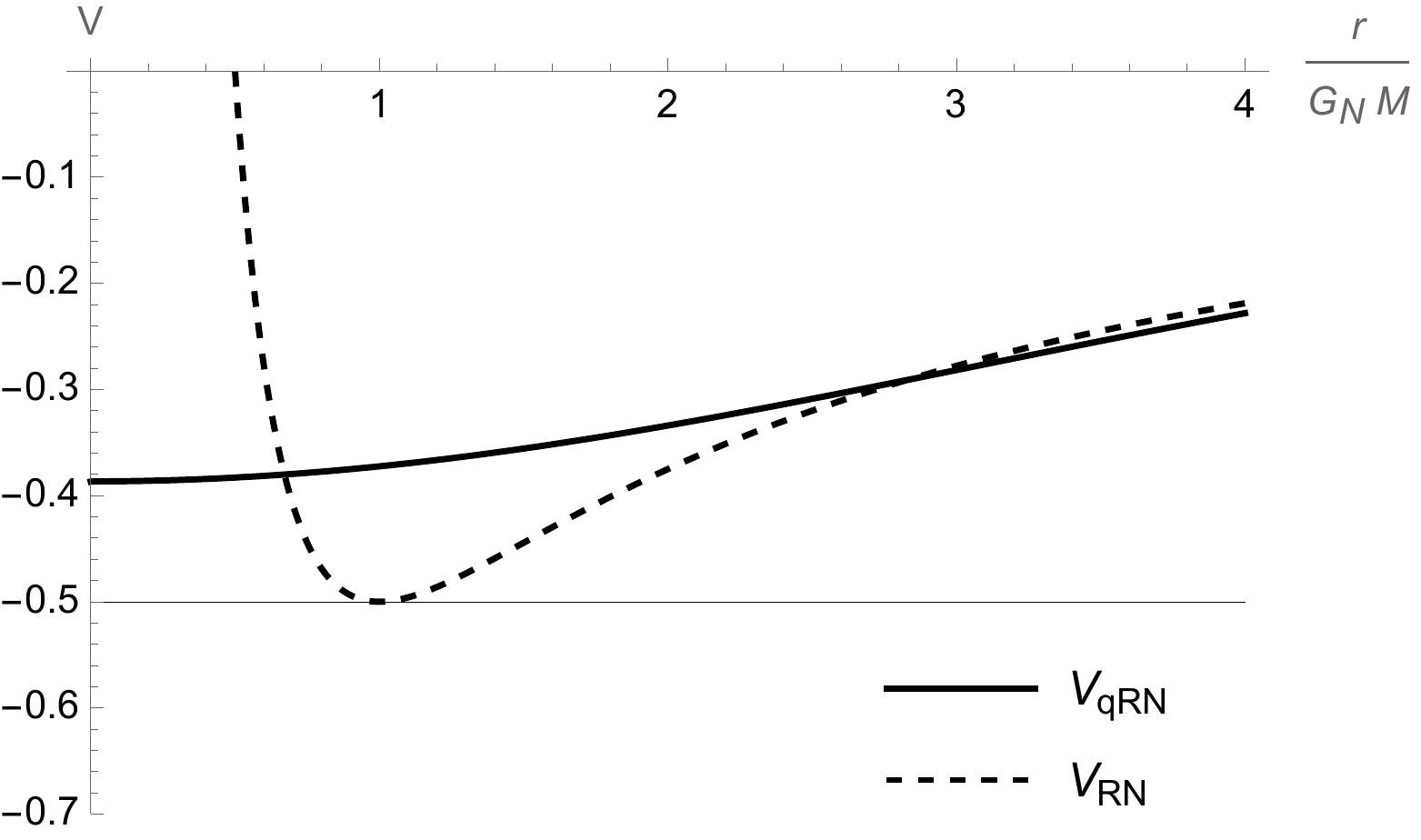}
$\ $
\includegraphics[width=8cm]{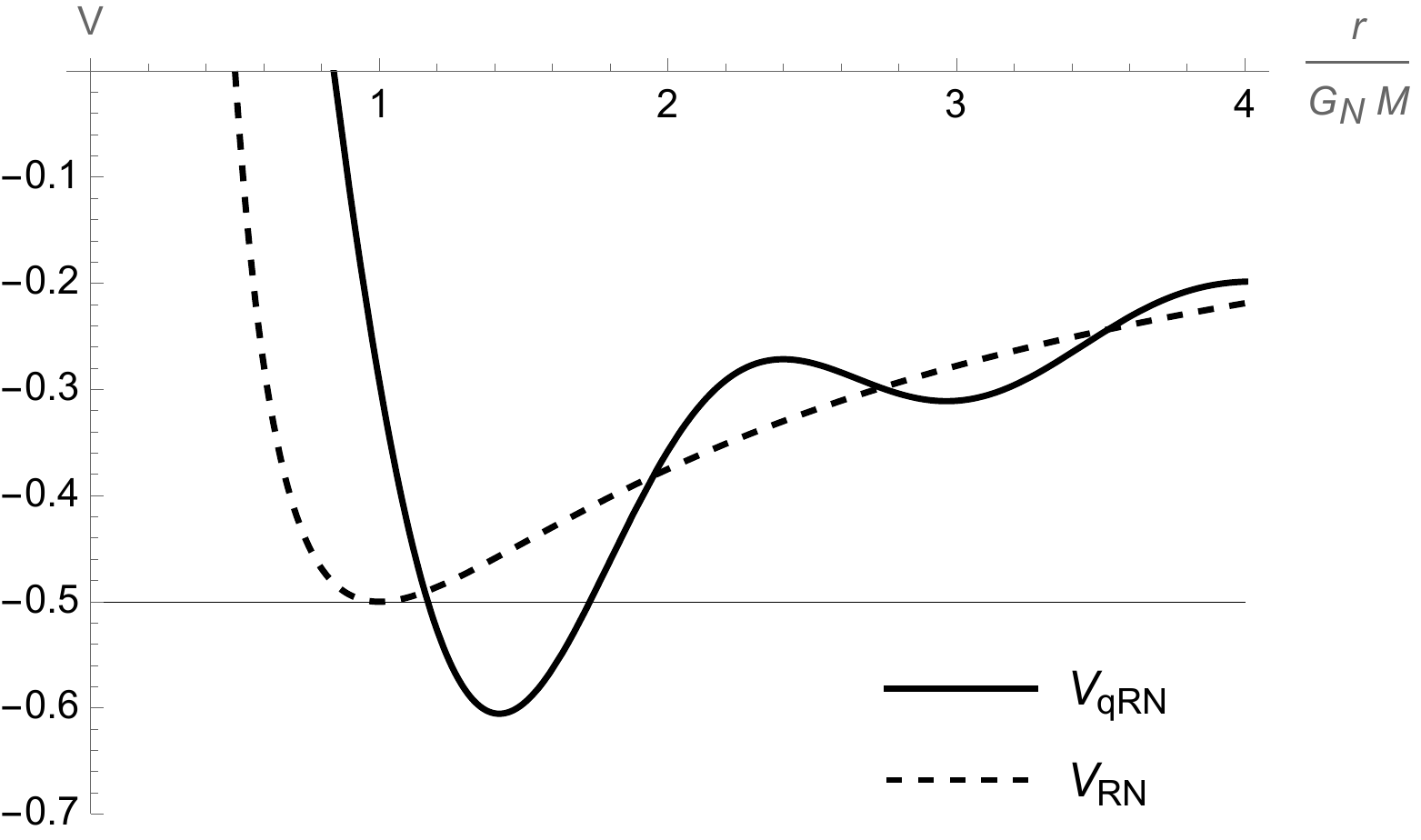}
\caption{Quantum potential $V_{\rm qRN}$ in Eq.~\eqref{VqRN} (solid line) compared to $V_{\rm RN}$
(dashed line) for the extremal case $R_-=R_+$ with $R_{\rm UV}=R_+$ (left panel) and $R_{\rm UV}=R_+/4$
(right panel).
The thin solid line $V=-1/2$ crosses the potential at the horizon.}
\label{chE}
\end{figure}

The quantum corrected causal structure for $R_-\lesssim R_{\rm UV}\lesssim R_+$ is in qualitative agreement
with the quantum mechanical description of the gravitational radius in Ref.~\cite{Casadio:2015rwa},
where the probability of finding the matter source inside the inner Cauchy horizon $R_-$ was shown to be 
small for masses above the Planck scale and charge $Q$ sufficiently below extremality.
One can then consider values of $M$ and $Q$ near the classical extremal case $R_+=R_-$, that is 
$Q^2\simeq \gn\,M^2$~\cite{Casadio:2015sda}.
Fig.~\ref{chE} shows two examples of extremal geometries for different values of the scale $R_{\rm UV}$.
For $R_{\rm UV}\gtrsim R_+$, the geometry is regular everywhere, with no horizons and no singularity.
If one lets $R_{\rm UV}$ fall below $R_+$, one in general obtains two (or more) horizons.
For a given $|Q|=\sqrt{\gn}\,M$, one can fine-tune $R_{\rm UV}$ so that
one degenerate horizon exists like in the classical case, but such a value can only be determined 
numerically (see Appendix~\ref{A:mass}).
\par
\begin{figure}[t]
\centering
\includegraphics[width=8cm]{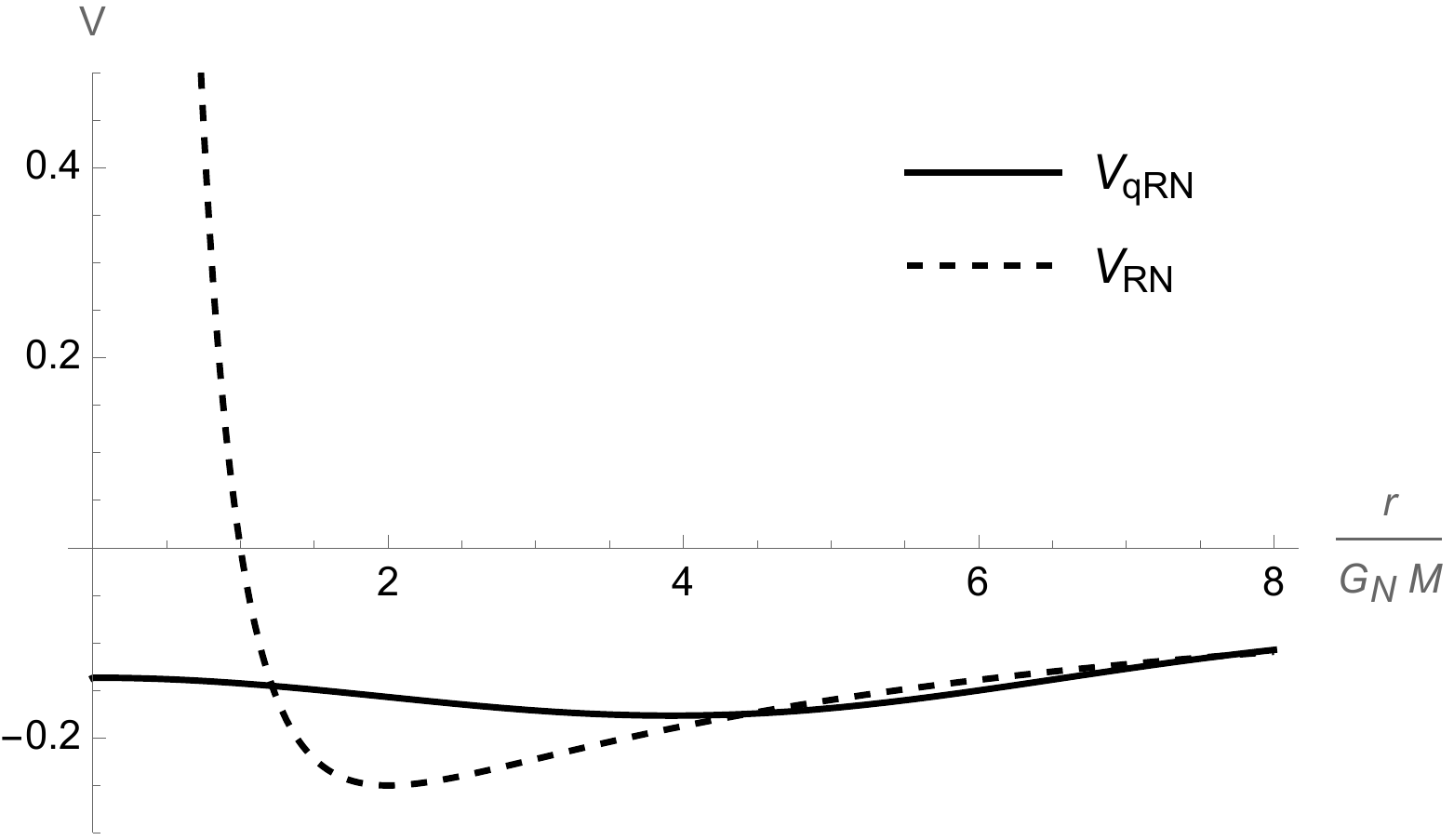}
$\ $
\includegraphics[width=8cm]{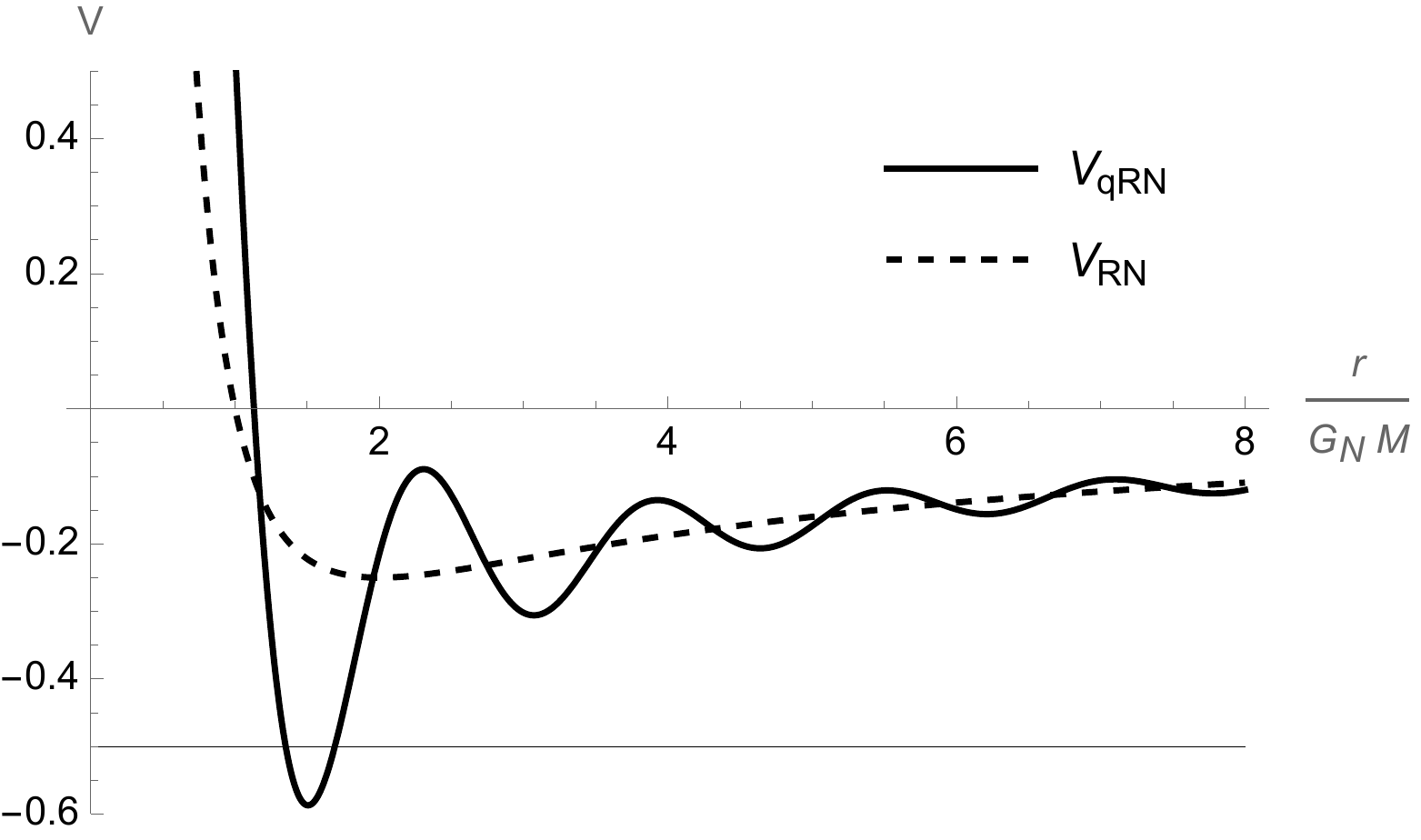}
\caption{Quantum potential $V_{\rm qRN}$ in Eq.~\eqref{VqRN} (solid line) compared to $V_{\rm RN}$
(dashed line) for classical naked singularity $\sqrt{\gn\,Q^2}=2\,\gn\,M$ with 
$R_{\rm UV}=\gn\,M$ (left panel) and $R_{\rm UV}=\gn\,M/4$ (right panel).
The thin solid line $V=-1/2$ crosses the potential at the horizon.}
\label{chN}
\end{figure}

For $Q^2> \gn\,M^2$, the classical Reissner-Nordstr\"om geometry becomes a naked singularity.
We show an example of the corresponding quantum corrected metric in Fig.~\ref{chN}.
For $R_{\rm UV}\gtrsim \gn\,M$, one again obtains a regular geometry, whereas for $R_{\rm UV}\lesssim \gn\,M$
a varying number of horizons in general reappears.
In this case, one can therefore have either a regular distribution of matter and charge, or a regular
black hole.
Like for the extremal case discussed above, the value of $R_{\rm UV}$
that separates the two different behaviors can only be determined numerically for given $M$ and $Q$.
\subsection{Thermodynamics}
\label{ss:th}
We recall that the Bekenstein-Hawking entropy is simply given by the area law~\cite{bekenstein}
\be
S_{\rm qRN}
=
\frac{\mathcal A_{\rm H}}{4\,\lp^2}
=
\frac{\pi\,R_{\rm q+}^2}{\lp^2}
\ee
and the black hole temperature~\cite{hawking}
\be
T_{\rm qRN}
=
\frac{\hbar\,\kappa_{\rm qRN}}{2\,\pi}
=
\frac{\hbar}{2\,\pi}\left.\frac{\d V_{\rm qRN}}{\d r}\right|_{r=R_{\rm q+}}
\ ,
\ee
where $\kappa$ is the surface gravity at the horizon.
Unfortunately, none of the above expressions can be computed analytically, because
$R_{\rm q+}$ can only be estimated numerically.
\par
For $M$ and $Q$ such that $R_-\lesssim R_{\rm UV}\lesssim R_+$, like the left panel in 
Fig.~\ref{ch}, deviations from the classical metric become very small and one therefore
expects just small numerical differences with respect to the classical expressions, that is
\be
S_{\rm qRN}
\simeq
S_{\rm RN}
=
\frac{\pi\,R_{+}^2}{\lp^2}
\ee
and
\be
\kappa_{\rm qRN}
\simeq
\kappa_{\rm RN}
=
\frac{\gn\left(M\,R_+-Q^2\right)}{R_+^3}
\ ,
\ee
where $R_+$ is given in Eq.~\eqref{R+-}.
\par
Much larger deviations are expected for $R_{\rm UV}\simeq R_+$, for which the
quantum corrected event horizon become much smaller then $R_+$ or disappears
(like in the left panel of Fig.~\ref{chX}).
\section{Conclusions and outlook} 
\label{s:conc}
\setcounter{equation}{0}
General relativity predicts the existence of singularities, which appear in black hole solutions
as the final product of the gravitational collapse.
This represents a clear limitation of the theory, and presumably its quantum version should
cure these pathologies.
Regular black holes represent simple workarounds to the problem of curvature singularity
that allow one to remain within the geometric description of general relativity, 
thus without resorting to any quantum argument.
However, these solutions often suffer of several caveats.
First, the matter distribution that generates these geometries has to violate the various
energy conditions that are typically ascribed to standard matter.
Nonetheless, such a scenario can be regarded as a mere effective description of a system
in a fully quantum regime, for which we still lack a proper UV description.
Second, and most importantly, such regular solutions typically entail the existence of inner
Cauchy horizons, signalling a breakdown of predictability.
\par
Starting from the idea that the classical geometry of a compact object should emerge
from a suitable description of the quantum state of both gravity and matter, we have 
reconstructed a quantum-corrected Reissner-Nordstr\"om geometry.
Such a geometry enjoys an integrable singularity, where tidal forces remain finite,
and the absence of inner Cauchy horizons when the UV cut-off $R_{\rm UV}$ is such that
$R_-\lesssim R_{\rm UV}\lesssim R_+$.
If the cut-off scale $R_{\rm UV}$ is associated with the final size of the collapsing object, it appears
sensible that it will never shrink below the would-be inner horizon, and the latter is therefore 
avoided.
\par
The same issues emerge in rotating black holes.
In order to provide a quantum description for these more complex classical geometries,
the approach from Ref.~\cite{Casadio:2021eio} employed here will have to be generalised,
for example by using a procedure similar to the one in Ref.~\cite{Contreras:2021yxe}. 
\section*{Acknowledgments}
R.C.~is partially supported by the INFN grant FLAG.
A.G.~is supported by the European Union's Horizon 2020 research and innovation
programme under the Marie Sklodowska-Curie Actions (grant agreement No.~895648). 
The work of R.C.~and A.G.~has also been carried out in
the framework of activities of the National Group of Mathematical Physics (GNFM, INdAM).
J.O.~is partially supported by ANID FONDECYT grant No.~1210041.
\appendix
\section{Mass function and horizon radius}
\label{A:mass}
\setcounter{equation}{0}
In the standard Reissner-Nordstr\"om metric, one can trade the dependence on the ADM mass $M$
for one of the zeros in Eq.~\eqref{R+-} of the metric functions $g^{rr}=-g_{tt}$ by defining
\be
f(r,Q)
=
\frac{\gn\,Q^2}{r^2}
\ ,
\ee
and
\be
g^{rr}
=
1-\frac{R_\pm}{r}\left[1+f(R_\pm,Q)\right]+f(r,Q)
=
1-\frac{R_\pm^2+\gn\,Q^2}{R_\pm\,r}
+\frac{\gn\,Q^2}{r^2}
\ .
\ee
The ADM mass is now given by
\be
2\,\gn\,M
=
R_\pm\left[1+f(R_\pm,Q)\right]
=
R_\pm
\left(
1+\frac{\gn\,Q^2}{R_\pm^2}
\right)
\ ,
\label{Mrh}
\ee
which yields Eq.~\eqref{R+-} as expected.
\par
For the quantum corrected metric~\eqref{gqRN}, it is easier to just solve Eq.~\eqref{Rq+-} for the mass $M$
as a function of a generic zero $r=\rh$, which yields
\be
2\,\gn\,M
=
\frac{\rh+\frac{\gn\,Q^2}{\rh}\left[1- \cos \left(\frac{\rh}{R_{\rm UV}}\right)\right]}
{\frac{2}{\pi}\,{\rm Si}\!\left(\frac{\rh}{R_{\rm UV}}\right)}
\ .
\label{Mh}
\ee
The metric function then reads
\be
g^{rr}
=
1
-
\frac{{\rm Si}\!\left(\frac{r}{R_{\rm UV}}\right)}
{{\rm Si}\!\left(\frac{\rh}{R_{\rm UV}}\right)}
\left\{
1
+
\frac{\gn\,Q^2}{\rh^2}
\left[1-\cos \left(\frac{\rh}{R_{\rm UV}}\right)\right]
\right\}
\frac{\rh}{r}
+
\frac{\gn\,Q^2}{r^2}
\left[1-\cos \left(\frac{r}{R_{\rm UV}}\right)\right]
\ .
\ee
By studying the function~\eqref{Mh}, one can in principle see for what values of $M$ and $Q$
there exists more values of $\rh$ depending on $R_{\rm UV}$.
In practice, this analysis can only be performed numerically.
\par
\begin{figure}[t]
\centering
\includegraphics[width=8cm]{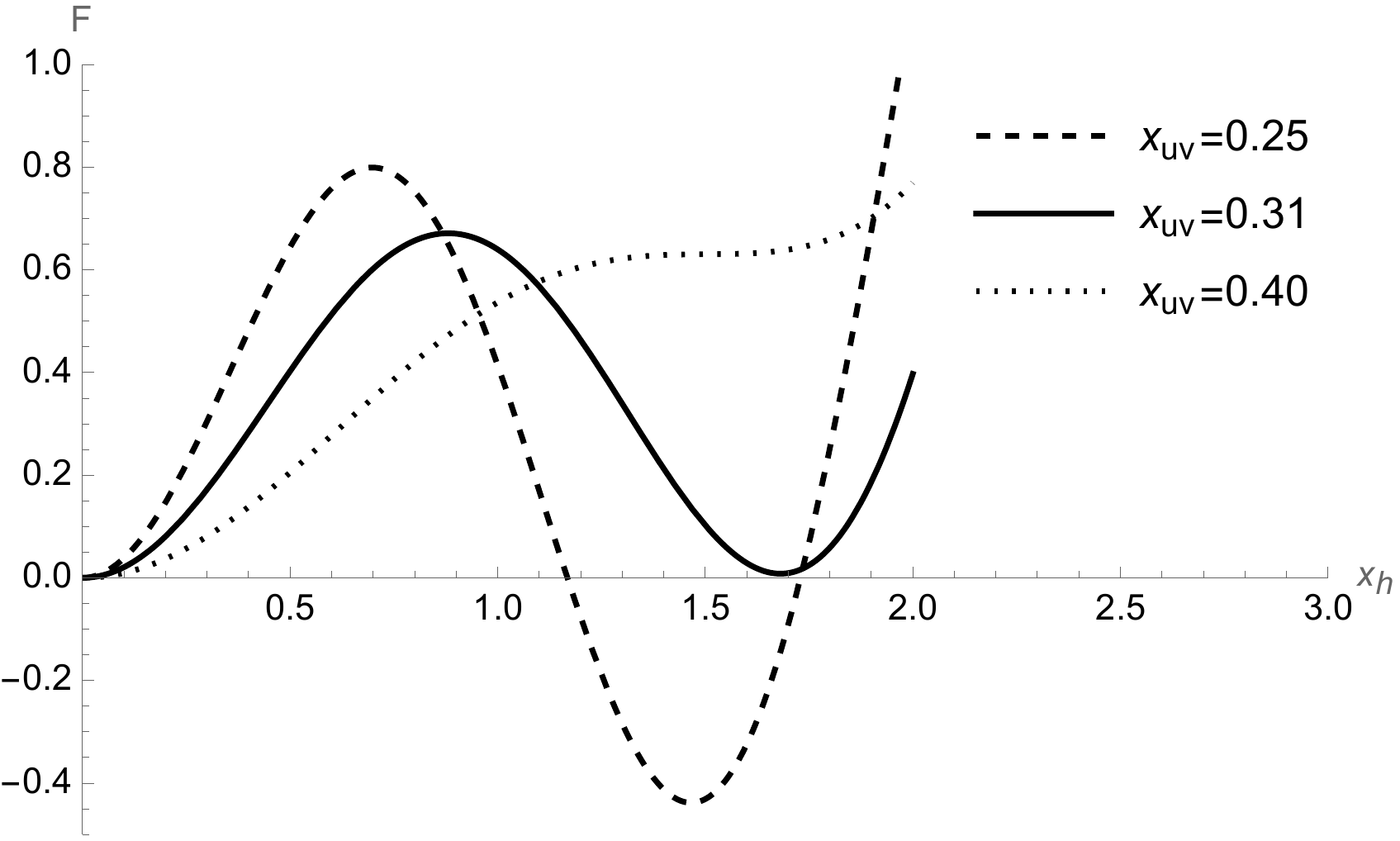}
\caption{The function $F$ in Eq.~\eqref{Fh} shows the existence of two horizons for $x_{\rm s}\simeq 0.25$
(dashed line), one degenerate horizon for $x_{\rm s}\simeq 0.31$ (solid line) and no horizons
for $x_{\rm s}\simeq 0.40$ (dotted line).}
\label{x}
\end{figure}

For example, in the classical extremal case $Q^2=\gn\,M^2$, Eq.~\eqref{Mrh} simplifies to
\be
F(x_{\rm h};x_{\rm s})
\equiv
x_{\rm h}^2
-\frac{4}{\pi}\,x_{\rm h}\,{\rm Si}\!\left(\frac{x_{\rm h}}{x_{\rm s}}\right)
+1
-
\cos \left(\frac{x_{\rm h}}{x_{\rm s}}\right)
=
0
\ ,
\label{Fh}
\ee
where we defined $x_{\rm h}\equiv \rh/\gn\,M$ and $x_{\rm uv}\equiv R_{\rm UV}/\gn\,M$.
Fig.~\ref{x} shows the function $F$ for three different values of $x_{\rm s}$ corresponding
to geometries with two horizons, one degenerate horizon and no horizon, respectively.  
\end{document}